\documentclass[journal]{vgtc}                
\ifpdf
\pdfoutput=1\relax                   
\usepackage{graphicx}                
\DeclareGraphicsExtensions{.pdf,.png,.jpg,.jpeg} 
\else
\usepackage{graphicx}                
\DeclareGraphicsExtensions{.eps}     
\fi%

\graphicspath{{figures/}{pictures/}{images/}{./}} 

\usepackage{microtype}                 
\PassOptionsToPackage{warn}{textcomp}  
\usepackage{textcomp}                  
\usepackage{mathptmx}                  
\usepackage{times}                     
\usepackage{cite}                      
\usepackage{tabu}                      
\usepackage{booktabs}                  
\usepackage{algorithm}
\usepackage[noend]{algpseudocode}
\usepackage[caption=false]{subfig} 
\usepackage{color}
\usepackage{multirow}
\usepackage{amsmath}

\makeatletter
\def\BState{\State\hskip-\ALG@thistlm}
\makeatother

\newcommand{\unsure}[1]{{#1}}
\newcommand{\rev}[1]{{#1}}
\newcommand*{\rot}[2]{
	\rotatebox[origin=l]{#1}{#2}
}

\onlineid{1140}

\vgtccategory{Research}
\vgtcpapertype{algorithm/technique}

\title{Data-Driven Space-Filling Curves}

\author{Liang Zhou, Chris R. Johnson, and Daniel Weiskopf}
\authorfooter{
	\item
	Liang Zhou and Chris R. Johnson are with SCI Institute, University of Utah. E-mail: {lzhou; crj}@sci.utah.edu.
	\item
	Daniel Weiskopf is with Visualization Research Center (VISUS), University of Stuttgart. E-mail: weiskopf@visus.uni-stuttgart.de.
}

\shortauthortitle{Zhou \MakeLowercase{\textit{et al.}}: Data-Driven Space-Filling Curves}

\abstract{We propose a data-driven space-filling curve method for 2D and 3D visualization. Our flexible 
	curve traverses the data elements in the spatial domain in a way that the resulting linearization better preserves features in space compared to existing methods. We achieve such data coherency by calculating a Hamiltonian path that approximately minimizes an objective function that describes the similarity of data values and location coherency in a neighborhood. Our extended variant even supports multiscale data via quadtrees and octrees. 
	Our method is useful in many areas of visualization, including multivariate or comparative visualization, ensemble visualization of 2D and 3D data on regular grids, or multiscale visual analysis of particle simulations. The effectiveness of our method is evaluated with numerical comparisons to existing techniques and through examples of ensemble and multivariate datasets. 
} 


\CCScatlist{ 
	\CCScat{K.6.1}{Management of Computing and Information Systems}%
	{Project and People Management}{Life Cycle};
	\CCScat{K.7.m}{The Computing Profession}{Miscellaneous}{Ethics}
}

\teaser{
	\centering
	\includegraphics[width=\linewidth]{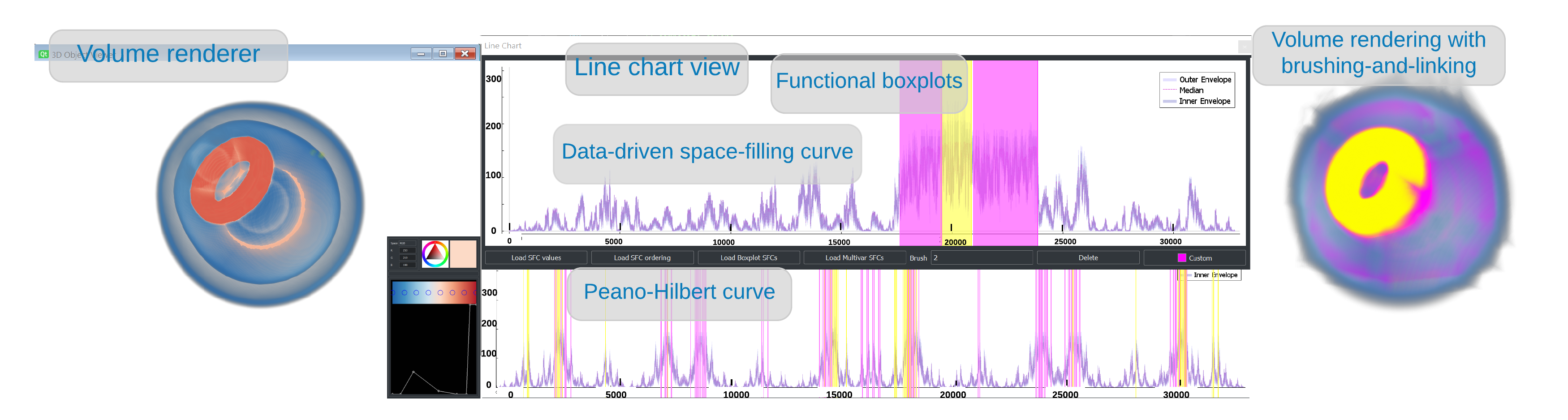}\vspace{-1em}
	\caption{Visualizations of an ensemble of volumetric data. The key idea is to employ a mapping from 3D space (as in the volume renderings) to 1D space via space-filling curves, which then allows us to show boxplots of the data distributions. 
		The ensemble is generated by sampling from Gaussian distributions of data values in the nucleon data with varying extents of uncertainty. The boxplots of the ensemble linearized with the Peano-Hilbert curve (bottom) do not preserve the coherency of 3D features---the small torus structure of high intensity cannot be readily identified. In contrast, our data-driven space-filling curve method (top) preserves features from 3D even in the 1D linearized representation as high intensities are more concentrated. 
		\rev{This observation is confirmed by brushing-and-linking---the torus could be covered by one brush and its surroundings with two brushes with our method (see the volume rendering on the right and the yellow and purple regions in ``Data-driven space-filling curve"), whereas multiple brushes are required by the Peano-Hilbert curve (yellow and purple regions in ``Peano-Hilbert curve").} }
	\label{fig:teaser}
}

\vgtcinsertpkg

\begin{document}
	
	
	\firstsection{Introduction}
	
	\maketitle
	Space-filling curves (SFCs) linearize an $n$-D image through a one-to-one mapping into one dimension.  
	Such linearization is useful in visualization as a tool for dimensionality reduction for 2D and 3D datasets. 
	In this paper, we propose a data-driven space-filling curve method for data on regular or multiscale grids.
	Our main goal is to preserve spatial coherency (i.e., locality) and data coherency (i.e., data features) at the same time. We construct a faithful representation of
	the original 3D or 2D data after linearization.
	The method is intended for easy feature identification in the 1D visualization of space-filling curves---as line-plots---and facilitates subsequent user interactions, e.g., brushing-and-linking in comparative visualizations.
	
	An important factor for choosing an appropriate space-filling curve is how well the locality of a dataset is preserved.
	Among typical space-filling curves, the Peano-Morton curve does not effectively preserve the locality, whereas the Peano-Hilbert curve~\cite{Hilbert:1891} is considered to have better locality. 
	Therefore, the Peano-Hilbert curve is popular in visualization.
	However, these space-filling curves ignore the content of a dataset.
	
	This issue is illustrated in~\autoref{fig:teaser}. 
	It shows the visualization of an ensemble of a nucleon volumetric dataset generated by sampling from Gaussian distributions with varying extents of uncertainty: the volume rendering generated with a blue-white-red color map is shown in \autoref{fig:teaser}~(left).
	Boxplots along the Peano-Hilbert curve are shown in \autoref{fig:teaser}~(central bottom), where the feature coherency in 3D is not preserved---the small torus structure of high intensity cannot be identified.  
	In fact, the torus is split into distant pieces in the 1D space and multiple brushes are required to select the feature (yellow areas in the ``Peano-Hilbert curve" of \autoref{fig:teaser}).
	By contrast, with our method (\autoref{fig:teaser}~(central top)), the torus can be identified as a feature spanning a much smaller range in 1D, and can be selected with a single brush (as seen in the yellow region) thanks to better preservation of features in the spatial domain.  
	With brushing-and-linking, the same regions are highlighted in yellow in 3D (\autoref{fig:teaser}~(right)) using linearizations with our method and the Peano-Hilbert curve. 
	\unsure{The better feature preservation of our method is also demonstrated with the purple brushes.}
	
	Our main contribution is a data-driven space-filling curve approach that comprises two variants of techniques: one for 2D and 3D regular grids, and another for 2D and 3D multiscale data.
	For regular grids, our method generates Hamiltonian cycles by replacing a minimum spanning tree using an objective function that combines locality and position terms; for multiscale data---quadtrees and octrees---our method finds adaptive Hamiltonian paths across data scales in a greedy fashion. 
	\rev{To enable the calculation of Hamiltonian paths for multiscale data, we make a second contribution: a simple and efficient technique that finds a Hamiltonian path given only the entry and exit edges (2D) and faces (3D) of bounding rectangles/boxes of (all vertices of) grid graphs (e.g., north, east, south, west of the bounding rectangle of a 2D grid graph).}
	We evaluate our method for each data type by comparing it to the Peano-Hilbert curve and scanline ordering on synthetic and real-world datasets.
	The effectiveness of our overall method is demonstrated through typical examples of 2D and 3D multivariate and ensemble data on regular grids and multiscale. 
	\rev{The source code of our method is available online\footnote{\url{https://github.com/zhou-l/DataDrivenSpaceFillCurve.git}}}.

	
	\section{Related Work}
	
	Space-filling curves~\cite{Sagan:1994:SFC}, discovered by Peano~\cite{Peano:1890}, are traditional topics in mathematics but now have various applications across different areas in computer science.
	Well-known space-filling curves include the Peano curve~\cite{Peano:1890}, the Gray code ordering~\cite{Faloutsos:TSE88}, and the Peano-Hilbert curve~\cite{Hilbert:1891}.
	These methods consider only spatial discretization on regular grids.  
	Adaptively refined space-filling curves are available for multiscale data structures, specifically, quadtrees and octrees, for dynamic load balancing for high-performance computing~\cite{campbell03a}.
	However, these methods use static configurations that are independent of the content of the data and relate only to the size of the data.

	The context-based space-filling curve~\cite{Dafner:cgf00} is one of the few examples of a data-dependent curve.
	It targets to improve autocorrelation in 2D image and video encoding. 
	\rev{There, a ``dual graph" (we use this redefinition by Dafner et al.~\cite{Dafner:cgf00} throughout our paper) is generated from the input image and then a minimum spanning tree of the graph is found, where weights are determined by an objective function.}
	Finally, the space-filling curve is constructed by replacing the minimum spanning tree with a Hamiltonian path from a Hamiltonian cycle.
	However, this method is limited to 2D data and does not support multiscale data, making it unsuitable for many visualization applications.
	Unlike this method, our data-driven space-filling curves support 3D volume data and multiscale data of 2D and 3D, which are not possible with the context-based space-filling curves~\cite{Dafner:cgf00}.
	In addition, our method introduces a new objective function that achieves both feature and locality coherency, making it more flexible than the context-based method.
	Another example is an approximation method of the space-filling curve with a data-driven metric~\cite{SKUBALSKARAFAJLOWICZ19971305}.
	However, only simple 2D examples with distributed points are demonstrated and it is unclear how the method could be extended to more complex data such as images and volumes. 
	A random space-filling curve method~\cite{Matias:spacefillcurve} based on the ``cover and merge" strategy is not data-driven but inspires the computational framework of the context-based space-filling curve~\cite{Dafner:cgf00} as well as our regular grid techniques. 
	
	Space-filling curves are useful for many visualization purposes. 
	They facilitate comparative visualizations due to locality preservation, i.e., points that are close on the space-filling curve are close in the original 2D/3D space (not necessarily the other way around). 
	Space-filling curves are used in ensemble visualization of 3D volumetric data~\cite{Demir:VIS14,Weissenboeck:vis18}.
	Peano-Hilbert curves are calculated for 3D ensemble data of multiple levels-of-details, and the linearized results are visualized as interactive enhanced line charts~\cite{Demir:VIS14} making comparisons of 3D members possible. 
	Similarly, a nonlinear compression method is available for the linearized 3D ensemble calculated using Peano-Hilbert curves~\cite{Weissenboeck:vis18}. 
	Hilbert attention maps~\cite{Netzel:etvis16} use Peano-Hilbert curves to visualize time-varying eye-tracking data sampled on 2D regular grids as a static image, allowing features of interest that span a small neighborhood to be traced easily in the attention maps.
	For all methods above, brushing-and-linking is used as the major exploration approach that relates the 1D linearization and the original data.
	Since our technique improves space-filling curves implementations for visualization, all of these visualization applications could potentially benefit from our method.
	
	Hamiltonian paths and cycles form the computational basis of our method.
	A Hamiltonian path/cycle is a path/cycle that visits each node in a graph exactly once, and a Hamiltonian cycle can be easily converted to a Hamiltonian path by a single cut on the cycle. 
	\rev{The computation of the general Hamiltonian path problem is NP-hard~\cite{Bollobas:1979}. For restricted scenarios, however, more efficient solutions are possible.}
	The existence/nonexistence of a Hamiltonian path is proven for 2D grid graphs~\cite{Itai:JOC1982}; for 3D graphs of even-numbered nodes along each dimension, a Hamiltonian path can be generated from a Hamiltonian cycle~\cite{Briais:CHES2012}.
	However, these methods require specified entry and exit nodes, which is infeasible for data-driven space-filling curves for multiscale data.  
	This is because if a path leaves a block of finer nodes and enters to a block of coarser nodes, we only know the exiting face of the block of finer nodes and the entering face of the block of coarser nodes.  
	\rev{We propose a more flexible Hamiltonian path generation method---for both 2D and 3D regular grids, given only edges/faces of entry and exit of a bounding rectangle/box---as a building block for our method.}

	Ensemble visualization, an active and challenging visualization topic~\cite{DBLP:journals/cga/ObermaierJ14}, is one of the target applications of our technique.
	Besides the aforementioned methods using space-filling curves~\cite{Demir:VIS14,Weissenboeck:vis18}, there are alternative techniques that use depth-based statistics~\cite{SCI:Gen2014a,SCI:Mir2014a,SCI:Mir14b,Raj:cga16}, scatterplots and parallel coordinates~\cite{SCI:Ros2016a}, trend graphs and parallel coordinates~\cite{DBLP:journals/tvcg/ObermaierBJ16}, and a flexible linked-view system with a configurable collection of statistical representations~\cite{SCI:Pot2009b}.
	Depth-based statistics is a fundamental building block for ensemble visualization. The computation and visualization of depth-based statistics is available for 1D functions~\cite{Sun:boxplots}, 2D surfaces~\cite{SCI:Gen2014a}, 2D contours~\cite{SCI:Mir2014a}, 3D contours~\cite{Raj:cga16}, and 2D and 3D curves~\cite{SCI:Mir14b}.
	In our paper, we employ a 3D extension of the surface boxplot~\cite{SCI:Gen2014a} together with our data-driven curves to visualize ensemble datasets.

	\section{Problem Formulation}
	\label{sec:method}

\begin{figure}[htb]
		\centering
		\includegraphics[width = 0.8\linewidth]{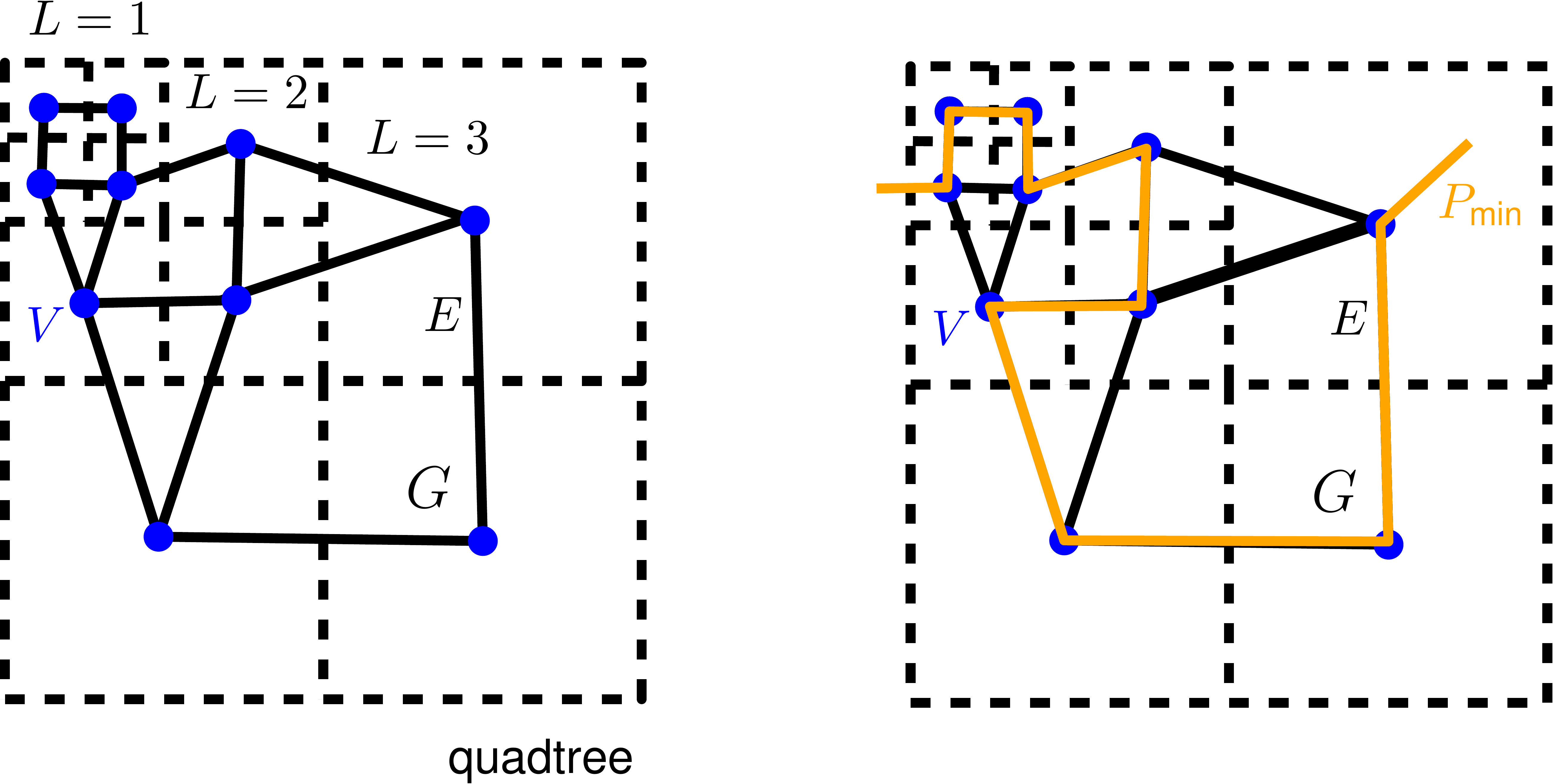}\\\vspace{-1em}
		\caption{\label{fig:generalProbMulti}A 2D multiscale graph $G = (V,E,L)$ on a quadtree with a Hamiltonian path $P_{\text{min}}$.}
	\end{figure}

	To support regular grids data and multiscale data with a unified representation, we model the input data in 2D and 3D as a graph:
	\begin{equation}
	G = (V, E, L)\;,\nonumber
	\end{equation}
	\rev{where vertices $V$ are nodes/vertices of the grid, edges $E$ connect neighboring vertices (typically 4-neighbor and 6-neighbor for 2D and 3D data, respectively), and $L$ is the level of the scale of the vertex. }
	Our formulation facilitates a flexible multiscale representation with the per-vertex scale $L$, as shown in~\autoref{fig:generalProbMulti}. 

\begin{figure}[htb]
		\centering
		\includegraphics[width= 0.9\linewidth]{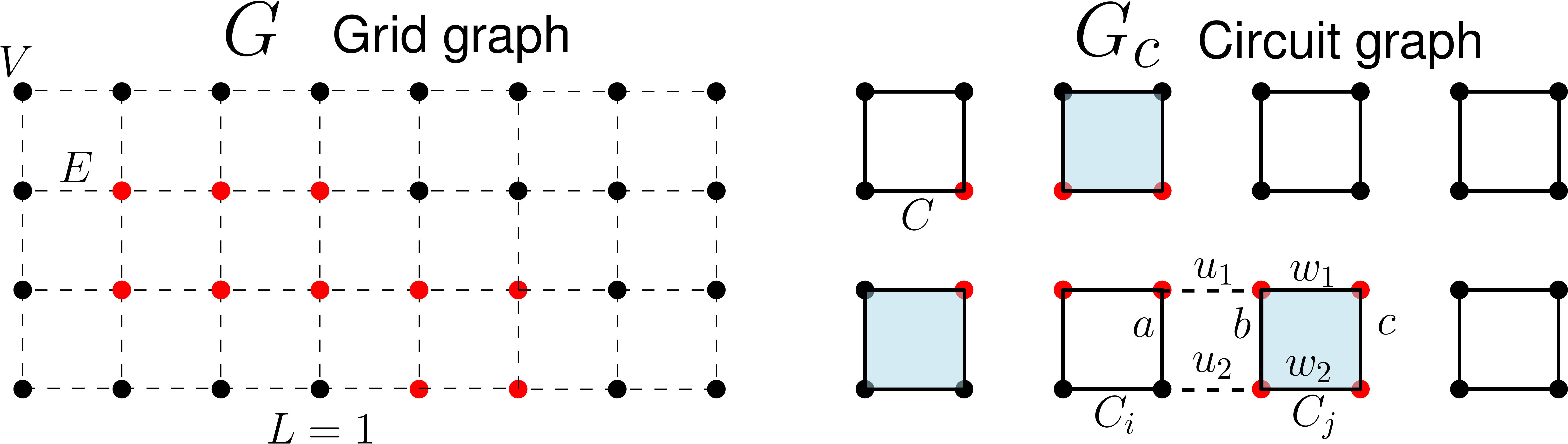}
		\caption{\label{fig:generalProbRegGrid}\rev{(Left) A 2D graph on a regular grid $G = (V,E,1)$ with corresponding data values (black = 0, red = 1), and its associated circuit graph $G_c$ (right) of circuits $C$. Adjacent circuits of $C_i$ are drawn in light blue. The edge weights of data values between circuits $C_i$ and $C_j$ are shown on the right.}}
	\end{figure}

	Regular grids are a special case of $G$ where the level is constant ($L = 1$) for all vertices and the graph becomes a grid graph (\autoref{fig:generalProbRegGrid}).
	
	\begin{figure*}[htb]
		\centering
		\includegraphics[width=0.9\linewidth]{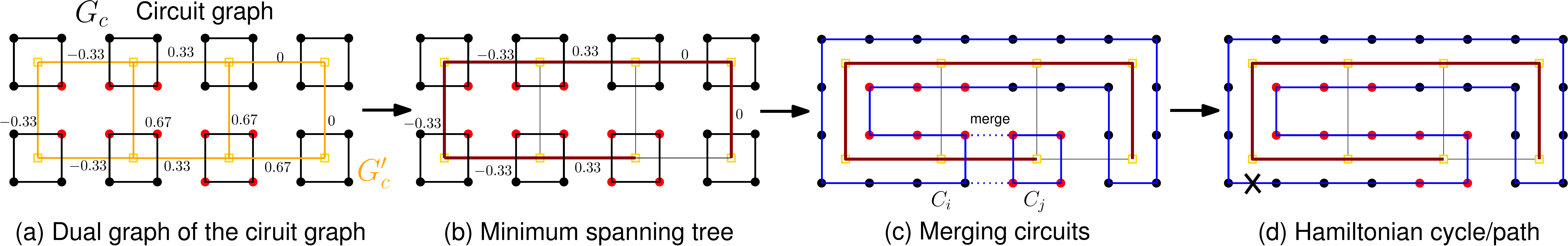}\vspace{-1em}
		\caption{Intermediate steps of the framework of~\cite{Dafner:cgf00} on 2D regular grids. The dual graph $G_c'$ (a) of circuits graph $G_c$ is a directed graph with our new weights (labeled on edges). Then, the minimum spanning tree (b) is found on the dual graph. Next, the Hamiltonian cycle is generated by merging circuits using the minimum spanning tree (c). Finally, (d) the Hamiltonian cycle is converted into a Hamiltonian path by making a cut anywhere on the cycle (shown as X). }\vspace{-1.5em}
		\label{fig:regGrid2DMinCase}
	\end{figure*}
	
	Typical space-filling curves~\cite{Peano:1890,Hilbert:1891,Sagan:1994:SFC} focus on the geometry of the curves, i.e., the geometry of $V$ and  $E$, which concerns only the preservation of locality but not data features as the curves are ignorant of the underlying data $s(V)$.
	Our goal is to generate space-filling curves that preserve both locality and data features. 
	This requires the scan order to be updated according to the data. 
	\rev{We focus on data-driven space-filling curves that traverse through connected nodes within the graph $G$.
		Counterexamples (curves that jump between unconnected nodes) are known to have poor locality coherency, e.g., the scanline and the Peano-Morton curve.
	}
	\rev{In our case, the space-filling curve problem is equivalent to a Hamiltonian path problem~\cite{Itai:JOC1982} with coherency preservation, which allows us to formulate the generation of data-driven space-filling curves as an optimization problem of Hamiltonian paths with an objective function that takes measures of both locality preservation and data-feature preservation into account.}
	\rev{Finding the minimum total weight of all possible Hamiltonian paths is hard on regular grids~\cite{Dafner:cgf00} and also on multiscale grids.}
	
	We denote a Hamiltonian path $P$ through all vertices $V$ as a sequence: 
	\begin{equation*}
	P = (v_1, v_2, \ldots, v_{n})\nonumber\;,
	\end{equation*}
	where  $v_i \in V$ is adjacent to $v_{i+1}$ for $1 \leq i < n$.
	We aim to find a path $P_{\text{min}}$ that minimizes an objective function $f(P)$:
	\begin{equation*}
	P_{\text{min}} = \arg\min_{P} f(P)\;.\nonumber
	\end{equation*}
	The objective function is formulated to be the sum of weights $W$ that is comprised of a feature preservation term $N$ that concerns data values $s(v)$ of vertex $v$, and a locality preservation term $R$:
	\begin{align}
	\label{eqn:objFunc}
	\centering
	f(P) &= \sum_{i=1}^{n-1}W(v_i, v_{i+1})\;,\\
	W(v_i, v_{i+1}) &= (1 - \alpha)N(s(v_i), s(v_{i+1})) + \alpha R(v_i, v_{i+1})\;,\nonumber
	\end{align}
	where $\alpha \in[0,1]$ is a user-set blend factor.
	\rev{Our locality preservation term is a simplified, first-order locality measure. The true locality measure of space-filling curves is multiscale, and, therefore, much more complicated. However, our simplified model still yields better positional coherency compared to the scanline and the context-based method, as shown in Section~\ref{sec:eval}.}
	
	\rev{Solving the minimization problem is infeasible except for extremely small datasets, and, therefore, we find an approximate optimum of the objective function.}
	\rev{For regular grids, the optimum of $f(P)$ is approximated by adopting the strategy used by the context-based method~\cite{Dafner:cgf00} but with our new objective function and an extension to 3D.
		Steps involved in the framework of regular grids are illustrated in \autoref{fig:regGrid2DMinCase}.
		
	}
	\rev{In Section~\ref{sec:regGrid}, we briefly review the setup of the framework of the regular grid and elaborate on our new objective function and its impact.
		The rationale and details of this framework can be found elsewhere~\cite{Dafner:cgf00}.
		For multiscale data, we propose approximately minimizing the objective function $f(P)$ using a top-down and recursive greedy algorithm, which is explained in Section~\ref{sec:multiscale}.}
	
	\section{Space-Filling Curve Generation for Regular Grids}
	\label{sec:regGrid}
	The steps for computing data-driven space-filling curves on regular grids are described in Algorithm~\ref{alg:2DregGrids}.
	\rev{Our new contributions are a new objective function as explained in Section~\ref{sec:regGridObjFunc} and the extension to 3D detailed in Section~\ref{sec:regGrid3D}.}
	
	\rev{We briefly review the computational framework~\cite{Dafner:cgf00} using a 2D example as illustrated in~\autoref{fig:regGrid2DMinCase}. }
	\rev{For a regular grid $G$ with an even number of vertices in each dimension, we first convert it to a graph $G_c$ of small circuits $C$ (\autoref{fig:generalProbRegGrid}), and then compute the dual graph $G_c'$ (refer to the redefinition in the context-based method~\cite{Dafner:cgf00}) of $G_c$ (\autoref{fig:regGrid2DMinCase}~(a)).
		With the dual graph of small circuits, we are able to find $P_{\text{min}}$ by constructing the minimum spanning tree of $G_c'$.
		The width and height of $G_c'$ are $w_d$ and $h_d$ respectively; each node of $G_c'$ corresponds to a circuit $C_k$ for $k\in \{1,\ldots,w_d\times h_d\}$ of 2$\times$2 vertices.
		The task of evaluating the weight between any vertex $v_j$ adjacent to vertex $v_i$ in $P$ is now transformed to evaluating the weight on circuits $W(C_i,C_j)$, where $C_i$ and $C_j$ are adjacent, and the dual of $C_i$ is already in the minimum spanning tree (\autoref{fig:generalProbRegGrid}~(right)).
		A minimum spanning tree is the tree that minimizes the sum of weights among all possible trees~\cite{Sedgewick:algorithm}, i.e., it guarantees to find $W(C_i,C_{i+1})$ as the minimum among all $W(C_i, C_j)$ in each step.
		The minimum spanning tree is built by joining edges of $G_c'$ using Prim's algorithm (\autoref{fig:regGrid2DMinCase}~(b)).
		Next, the minimum spanning tree is converted to a Hamiltonian cycle by merging the circuits according to the minimum spanning tree with the cover-and-merge strategy~\cite{Matias:spacefillcurve} (\autoref{fig:regGrid2DMinCase}~(c)).
		Finally, a Hamiltonian path $P_{\text{min}}$ is created by making a single cut anywhere in the Hamiltonian cycle~\cite{Dafner:cgf00} (\autoref{fig:regGrid2DMinCase}~(d)). }
	
	\begin{algorithm}[htb]
		\caption{Data-Driven SFC for Regular Grids}\label{alg:2DregGrids}
		\begin{algorithmic}[1]
			\Procedure{ddSFCRegGrid}{$G$}
			\State $G_c' \gets$  \textbf{buildSmallCircsDualGraph}($G$) \Comment{$G$---2D/3D grid graph, $G_c'$---dual graph of small circuits graph of $G$}
			\State $W \gets$ \textbf{calculateDualGraphWeights}($G_c'$) \Comment{$W$---weights on $G_c'$}
			\State MST  $\gets$ \textbf{findMinSpanTree}$(G_c', W)$ \Comment{MST---minimum spanning tree}
			\State $P_{\text{min}} \gets$ \textbf{mergeHamCycleAndCut}(MST, $v_s$)
			\Comment{$v_s$---entry vertex, $P_{\text{min}}$---data-driven SFC}
			\State \Return $P_{\text{min}}$ 
			\EndProcedure
		\end{algorithmic}
	\end{algorithm}

	\subsection{Objective Function}
	\label{sec:regGridObjFunc}
	\rev{In our new objective function, the definition of weights of circuits on regular grids by adding the circuit $C_j$ to the minimum spanning tree reads:}
	\begin{equation}
	\label{eqn:objFunc}
	W(C_i,C_j) = (1 - \alpha)N(C_i,C_j) + \alpha R(C_i, C_j)\;,\quad \alpha \in [0,1] \;,   
	\end{equation}
	\rev{where $C_i$ and $C_j$ are adjacent circuits, and the dual of $C_i$ is already in the minimum spanning tree.}
	
	\begin{figure}[b!]
		\vspace{-1em}
		\centering
		\includegraphics[width = 0.5\linewidth]{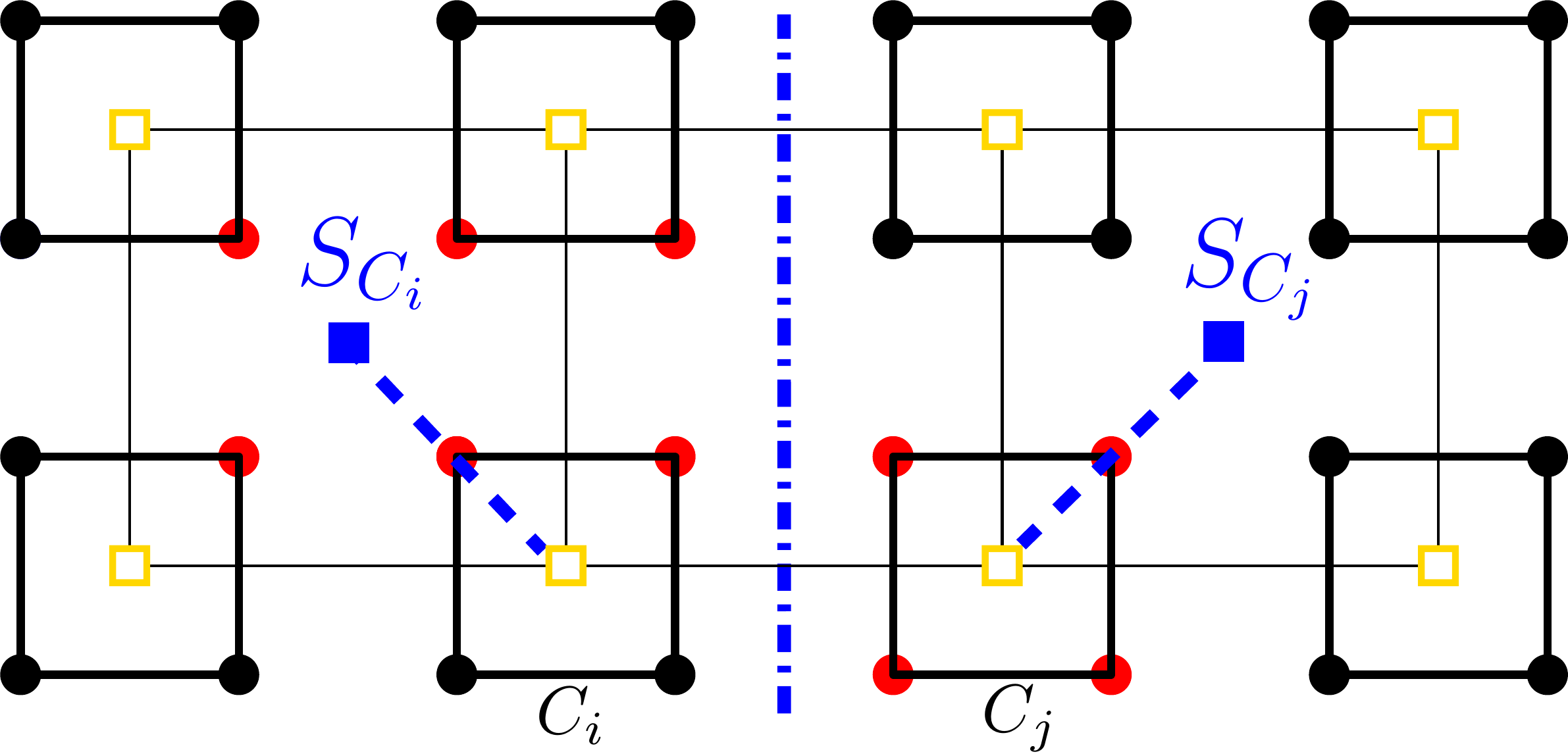}
		\caption{$G_c'$ is partitioned  (the blue dash-dot line) into blocks denoted by their centers (e.g., $S_{C_i}$ and $S_{C_j}$) to accommodate the new positional term of the objective function.} 	
		\label{fig:regGridBlock}
	\end{figure}
	
	\newcommand{\isabelHeight}{5cm}
	\begin{figure*}[htb]
		\centering
		\subfloat{\includegraphics[width=0.9\linewidth]{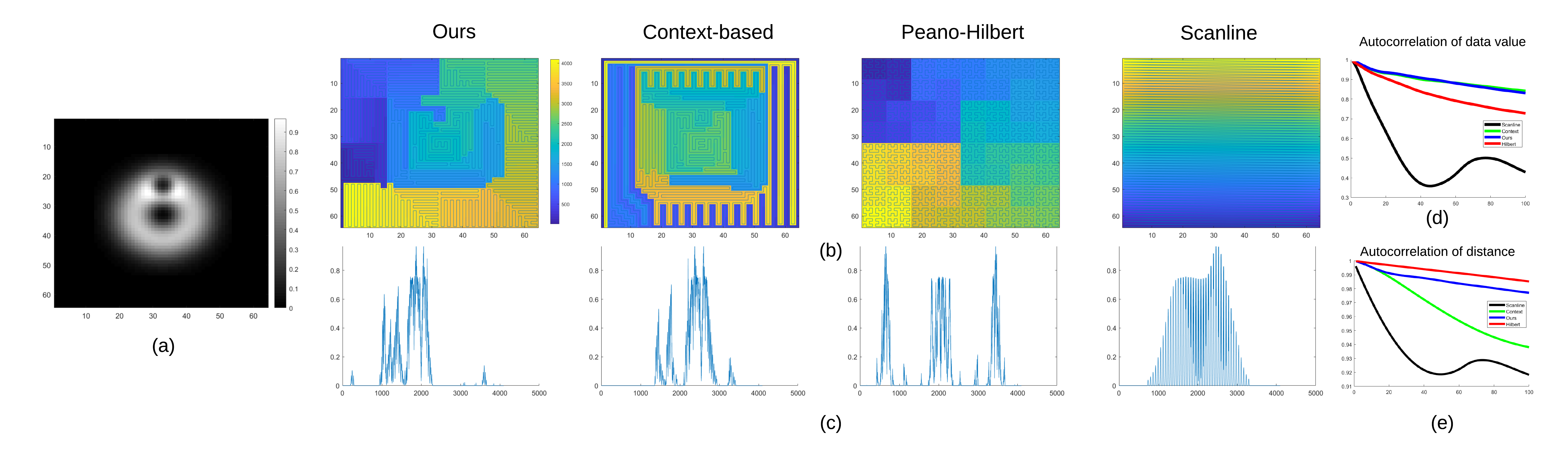}}\vspace{-1.5em}
		\caption{Comparison of linearization methods applied to a slice of the nucleon dataset (a). Our new data-driven space-filling curve is compared to previous methods: the context-based space-filling curve, the Peano-Hilbert curve, and  scanline ordering. The spatial layout of the respective curves is shown in (b), and the linearization of the data values in (c). 
			\rev{The spatial layout (b) is colored by the traversal order of curves (the horizontal axis of (c)) with the parula colormap (right of (b) Ours). }
			Autocorrelations of value (d) and radial distance (e) quantify data coherency and locality preservation, respectively. The plots show that our approach provides the best compromise between the two conflicting goals. Note that the autocorrelations of data values of our method and the context-based method are largely overlapping.}
		\label{fig:hurricane2D}
		
		\subfloat{\includegraphics[width = \linewidth]{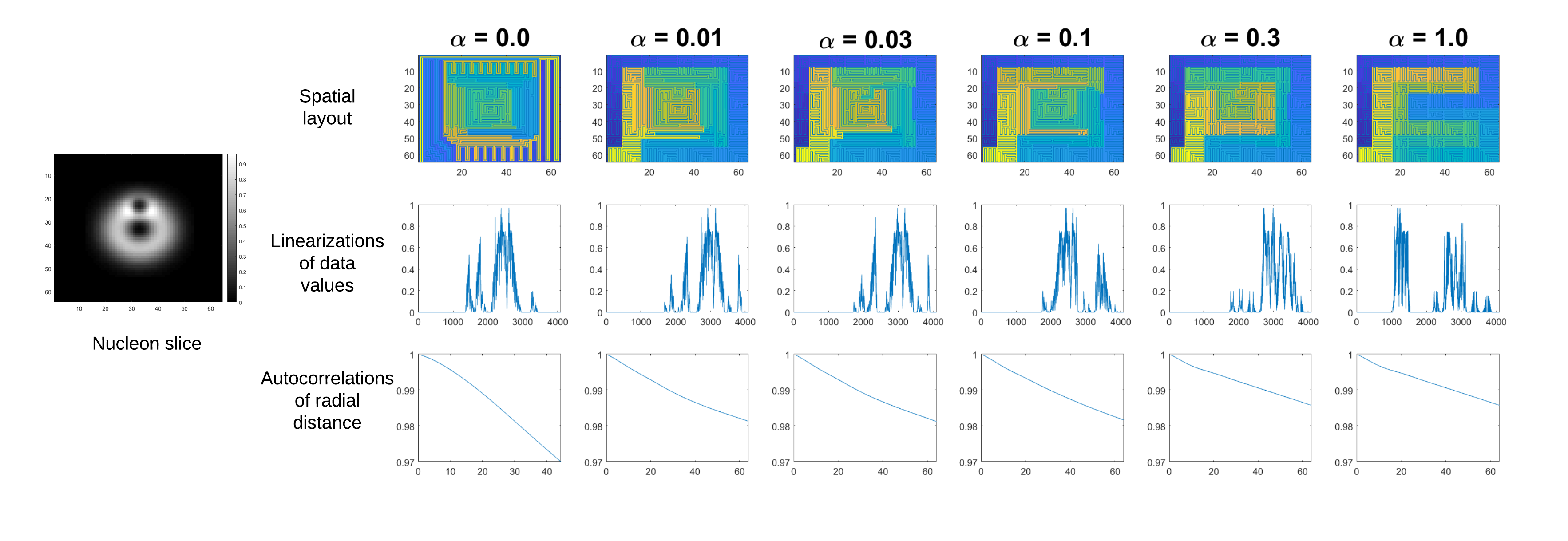}}\vspace{-2.5em}
		\caption{ The effect of different $\alpha$ (left--right: 0, 0.01, 0.03, 0.1, 0.3, and 1.0) on the scan order of our space-filling curves for the nucleon slice data (left). The order is color-coded using the same color map as in~\autoref{fig:hurricane2D}. }\vspace{-1em}
		\label{fig:alphaEffect}
	\end{figure*}
	
	\rev{For value coherency, we reuse the definition of value weights of circuits of the context-based method~\cite{Dafner:cgf00}. 
		The value weight that grows the minimum spanning tree with $C_j$ reads: }
	\begin{equation}
	\label{eqn:weight2D}
	N(C_i, C_j) = |u_1|+|u_2|+|w_1|+|w_2|+|c|-|b|-|a|\;,
	\end{equation}
	\rev{where $u_1$, $u_2$, $w_1$, and $w_2$ belong to edges along the growing direction of the minimum spanning tree, whereas $a$, $b$, and $c$ belong to faces across the growing direction. 
		All values above are differences of data values $s(v)$ of vertices along corresponding edges in the grid graph as defined in~\autoref{fig:generalProbRegGrid}~(right).}
	
	To measure the positional coherency, we first partition the dual graph into blocks with width $w_b$ and height $h_b$, and denote the block center of circuit $C_k$ as $S_{C_k}$, as shown in~\autoref{fig:regGridBlock}. 
	Then, we derive our positional coherency term that measures the distance of the 2D position of the circuit to the block center.
	The positional term is defined as follows:
	\begin{align}
	R(C_i, C_j) &= R_\text{pos}(C_j) = ||(C_j.x, C_j.y) - (S_{C_j}.x, S_{C_j}.y)||\;,
	\end{align}
	\rev{where $R_\text{pos}(C_j)$ measures the positional difference as the spatial distance between $C_j$ and the center of the block $S_{C_j}$.
		Since an edge weight is required for finding the minimum spanning tree, we assign $R_\text{pos}(C_j)$ to the edge $C_i$--$C_j$ in the dual graph to facilitate a unified weight definition with the value term. 
	}
	

	A comparison of our data-driven curve for 2D regular grids and other linearization techniques is shown in~\autoref{fig:hurricane2D}.
	It can be seen that our method (\autoref{fig:hurricane2D}~(Ours)) yields coherent results and correctly reveals the two peaks as coherent and neighboring features. 
	The context-based space-filling curve (\autoref{fig:hurricane2D}~(Context-based)) also reveals such structures but its spatial layout is not localized (\autoref{fig:hurricane2D}~(b, Context-based)), which is confirmed by a similar autocorrelation of value (\autoref{fig:hurricane2D}(d)) and a inferior autocorrelation of radial distance (\autoref{fig:hurricane2D}(e)) compared to our new method). 
	This indicates that our new method yields more coherent results than the context-based method.
	%
	The scanline order (\autoref{fig:hurricane2D}~(Scanline)) generates a cluttered line chart that goes up and down and it is not possible to see the data content; the Peano-Hilbert curve (\autoref{fig:hurricane2D}~(Peano-Hilbert)) fails to show the two bright regions as neighboring features, and the concentrated overall structure is shown along the whole span of the line chart.
	
	In our method, the blend factor $\alpha$ allows the user to flexibly change the importance of value coherency and positional coherency, which is not possible in the context-based space-filling curve~\cite{Dafner:cgf00}. 
	\autoref{fig:alphaEffect} shows the effect of $\alpha$ on the traversal order of an image. 
	\rev{The impact of $\alpha$ on value coherency and positional coherency is data-dependent and nonlinear.
		We empirically used an $\alpha$ value of 0.1 (except for~\autoref{fig:heart}, where $\alpha=0$) as tests on datasets for evaluation (Section~\ref{sec:eval}) show that such a value yields a good balance between the positional coherency and data-value coherency.
		We recommended using $\alpha=0.1$ as a default. Fine-tuning using trial-and-error may be required for a specific dataset to achieve desired properties.}

	\subsection{3D Volumes}
	\label{sec:regGrid3D}
	
\begin{figure}[tb]
		\centering
		\subfloat{\includegraphics[trim={1cm 1.8cm 1cm 0cm}, clip,width = \linewidth]{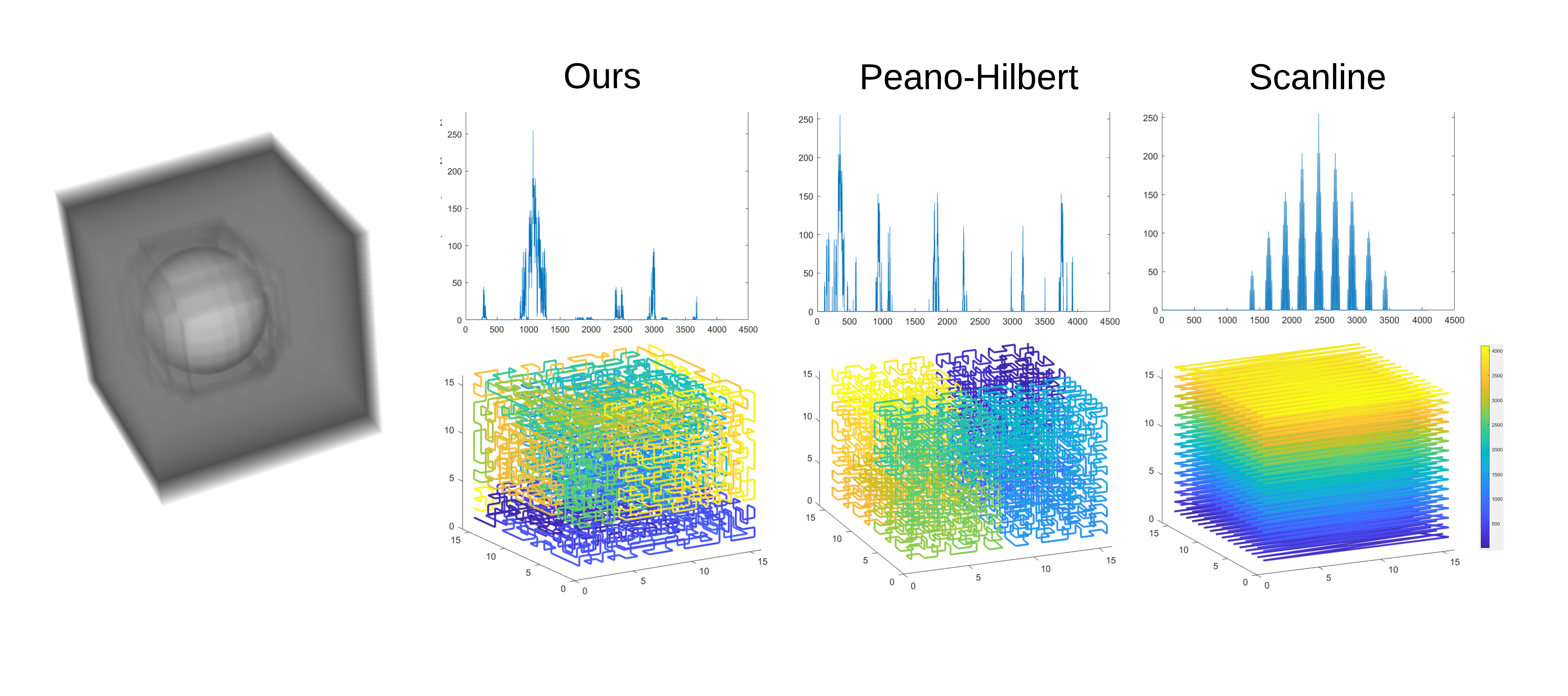}}\vspace{-1em}
		\caption{Linearizations (top) of a synthetic volume data of a sphere (left) with our data-driven curve, the Peano-Hilbert curve, and scanline ordering. The scan orders of curves are shown in the bottom row. }
		\label{fig:sfc3D}
		
		\centering
		\subfloat[]{\includegraphics[width = 0.64\linewidth]{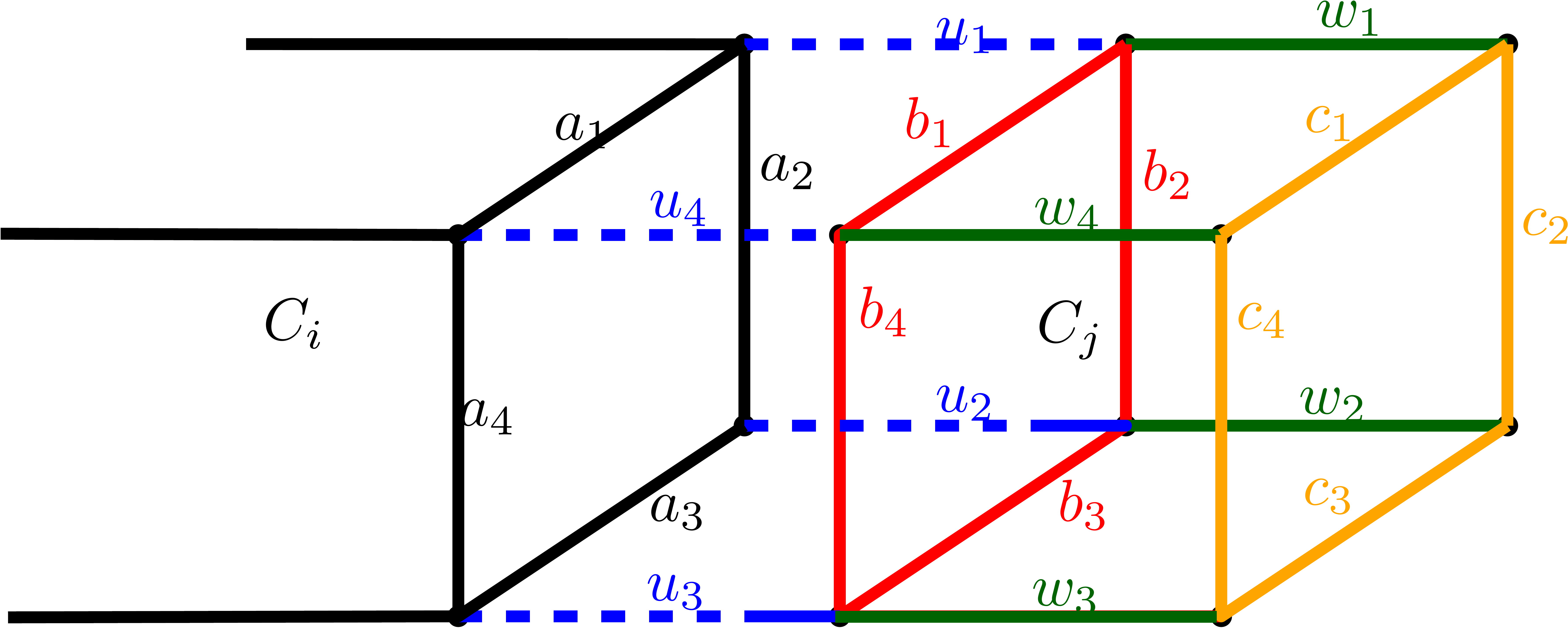}}\vspace{-1em}\\\hfill
		\subfloat[]{\includegraphics[width = 0.3\linewidth]{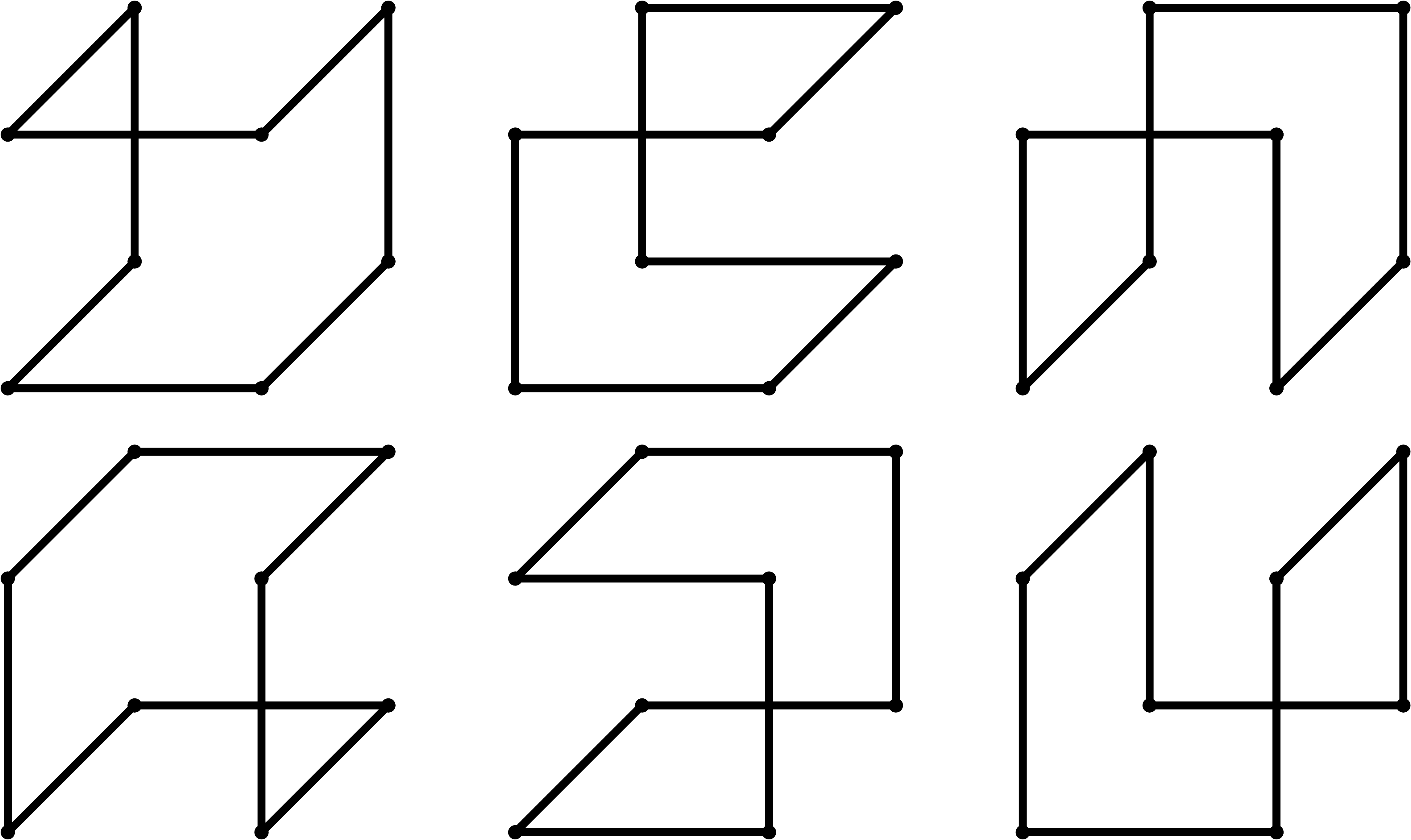}}\hfill
		\subfloat[]{\includegraphics[width = 0.64\linewidth]{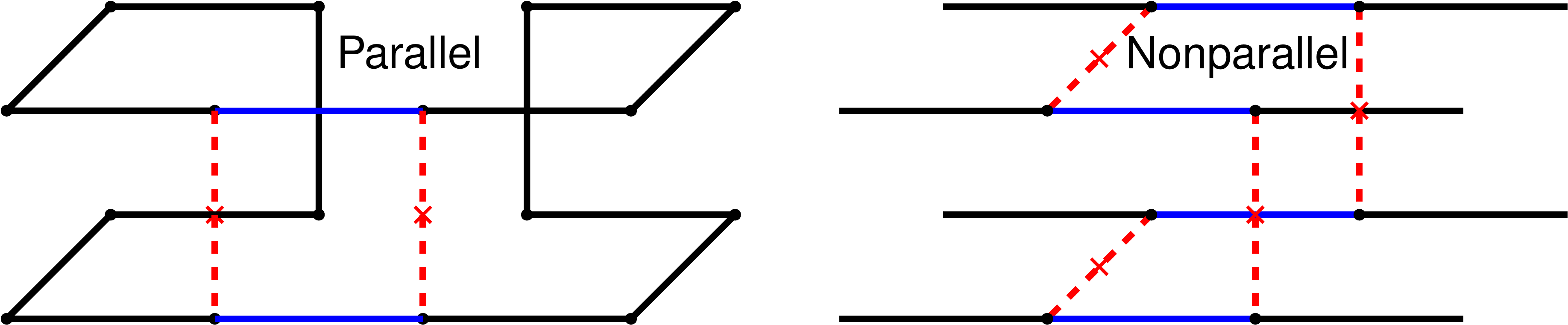}}
		\caption{The value weights of cubes on 3D regular grids (a). The cubes need to be converted into cycles during merging. There exist (b) six cycle configurations of a unit cube, and the cycles are merged with (c)~two association rules.}\vspace{-1em}
		\label{fig:3Dcycles}
	\end{figure}

	Because a data-driven or context-based space-filling curve technique for 3D data on regular grids is useful for visualization applications~\cite{Demir:VIS14,Weissenboeck:vis18}, we extend our data-driven space-filling curve to 3D regular grids.
	\autoref{fig:sfc3D} shows a comparison of linearizations of a synthetic volume data---a sphere with increasing data value from exterior to interior (\autoref{fig:sfc3D}~(d))---using our data-driven method, the Peano-Hilbert curve, and scanline ordering.
	It can be seen that our method (\autoref{fig:sfc3D}~(a)) best preserves the value signature of the sphere as a concentrated continuous single peak with least noise, which is not possible with the Peano-Hilbert curve (\autoref{fig:sfc3D}~(b)) or the scanline (\autoref{fig:sfc3D}~(c)), the sphere is split into many pieces make the feature unidentifiable.

	\begin{figure*}[tb]
		\centering
		\includegraphics[trim={1cm 1.8cm 1cm 0cm}, clip, width = 0.95\linewidth]{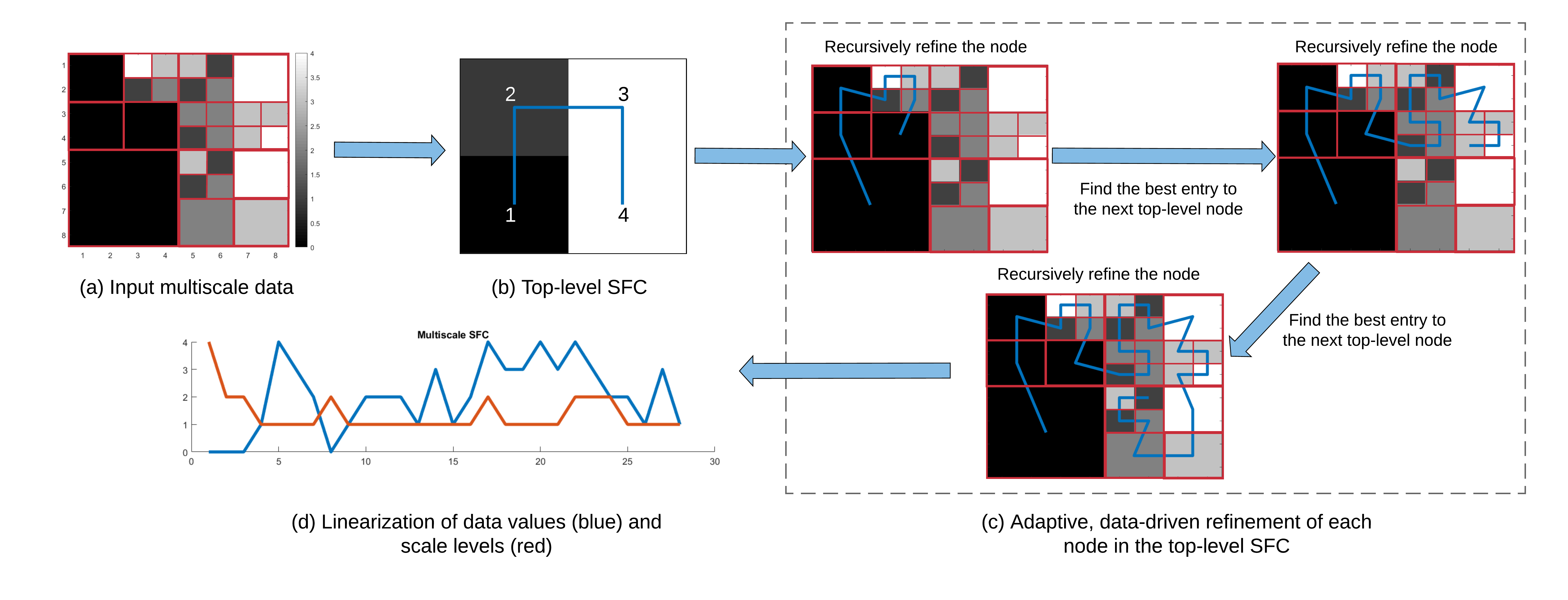}\vspace{-1em}
		\caption{Steps to compute a data-driven space-filling curve for multiscale data. }\vspace{-1em}
		\label{fig:multiscaleSteps}
	\end{figure*}
	
	We extend the computational framework of the aforementioned 2D method (Algorithm~\ref{alg:2DregGrids}) to 3D.
	\rev{Here, $G_c$---the equivalent to the circuit graph in 2D---is now comprised of small unit cubes---of 2$\times$2$\times$2 voxels---instead of circuits.}
	The weights in the objective function have the same form as of Equation~\ref{eqn:objFunc}, but the two coherency terms are modified accordingly for 3D:
	\begin{align}
	\vspace{-1em}
	N(C_i,C_j) &=\sum_{r=1}^{4}(|u_r| + |w_r| + |c_r| - |b_r| - |a_r|)\;,\nonumber\\
	R(C_i, C_j) &= R_\text{pos}(C_j) \;,\nonumber\\
	R_\text{pos}(C_j) &= ||(C_j.x,C_j.y,C_j.z)-(S_{C_j}.x,S_{C_j}.y,S_{C_j}.z)||\;.
	\label{eqn:reggridWeights3D}
	\end{align}
	\rev{As an analogy to the 2D case, $u_r$ and $w_r$ are edges along the growing direction, while $a_r$, $b_r$, and $c_r$ are faces across the growing direction. The value weights of cubes on 3D regular grids are illustrated in~\autoref{fig:3Dcycles}~(a).}
	Instead of four neighbors (top, down, left, right) in the 2D case, six neighbors (with front and back as additional neighbors) are used in the 3D case when building the minimum spanning tree. 
	
	The fact that a 3D unit cube is no longer directly a cycle as in the 2D case makes the conversion from the minimum spanning tree to a Hamiltonian cycle more complicated.
	There exist six possible cycle configurations in a unit cube as shown in~\autoref{fig:3Dcycles}~(b).
	
	\rev{After building the minimum spanning tree, we grow the Hamiltonian cycle by traversing the tree and associating unit cycles with a random configuration (from the six configurations) with association rules~\cite{Briais:CHES2012}. }
	Two association rules for two neighboring unit cycles are adopted (\autoref{fig:3Dcycles}~(c)): if parallel neighboring edges exist, we break the parallel edges and associate the four endpoints; if parallel edges do not exist, we need to break the neighboring edges and associate all eight endpoints.
	
	
	
	\unsure{Our data-driven technique could be extended for $n$ data attributes, where multidimensional data values live in regular grids in a 2D or 3D spatial domain.}
	The vertices in the graph now have vector-based data values, and the weights of the objective function can be defined as certain metrics of the vectors, e.g., L2-norm.
	\unsure{Our current visualization is based on the linearization of one data attribute for multidimensional data, and one member (typically, the median) for ensemble datasets.  }

	\section{Multiscale Data-Driven Space-Filling Curves}
	\label{sec:multiscale}
	Our data-driven curve for a multiscale data is equivalent to finding a minimum Hamiltonian path that traverses every leaf node in a quadtree or octree $T$.
	\rev{However, the aforementioned regular grid strategy is not applicable to build a dual graph for multiscale graphs because they often do not contain even-numbered vertices along each dimension (\autoref{fig:multiscaleSteps}~(a)), and, therefore, a Hamiltonian cycle does not exist. 
		As a result, we resort to an approximation strategy to the minimum Hamiltonian path on multiscale grids with top-down adaptive refinement.
	}
	
	
	\begin{algorithm}[htb]
		\caption{Data-Driven SFC for Multiscale Data}\label{alg:quadtree}
		\begin{algorithmic}[1]
			\Procedure{SFCmultiscale}{$\{I_1, I_2, \cdots, I_{L_c}\}$, $T$} 
			\Comment{T: tree structure (quadtree/octree)}
			\State  $P_{\text{top}} \gets$ \textbf{findTopLevelSFC}$(I_{L_c})$ \Comment{computes $P_{\text{top}}$---the top level SFC (data-driven)}
			\State $P_{\text{min}} \gets$ [] \Comment{$P_{\text{min}}$---SFC of the whole data}
			\State $v_{\text{last}} \gets \mathbf{0}$  \Comment{ $v_{\text{last}}$---the last SFC node}
			\For {$i$ in range(1, length($P_{\text{top}}$))} 
			\State block $\gets$  $T(P_{\text{top}}[i]$)) \Comment{retrieves the corresponding block of the current SFC node from the tree}
			\State $Pz \gets$ \textbf{refine}($\{I_1, I_2, \cdots, I_{L_c}\}$, $T$, block, $v_{\text{last}}$) \Comment {$Pz$---SFC of the current block computed by adaptive refinement (data-driven)}
			\State $P_{\text{min}} \gets [P_{\text{min}},Pz]$ \Comment{appends $P_{\text{min}}$ with $Pz$}
			\State $v_{\text{last}} \gets P_{\text{min}}[\text{last}]$ \Comment{records the last member of $P_{\text{min}}$}
			\EndFor
			\State \textbf{return} $P_{\text{min}}$
			\EndProcedure
		\end{algorithmic}
	\end{algorithm}
	The process of our mutliscale data-driven space-filling curve method is summarized in Algorithm~\ref{alg:quadtree}.
	Given a multiscale dataset on a quadtree (\autoref{fig:multiscaleSteps}~(a)) or octree whose nodes are of levels $1 \leq L \leq  L_c$, where 1 is for the finest level and $L_c$ is for the coarsest level, we prepare an image/volume pyramid of $L_c$ levels $\{I_1, I_2, \cdots, I_{L_c}\}$ for subsequent computations. 
	First, the top-level space-filling curve $P_{\text{top}}$ of the coarsest level $I_{L_c}$ is found (\autoref{fig:multiscaleSteps}~(b)).
	Based on the number of nodes in the coarsest pyramid level, the path is calculated using either the regular-grid-based data-driven curve method as described in Section~\ref{sec:regGrid} or the general Hamiltonian path method as discussed in Section~\ref{sec:hamPath}.
	Then, we adaptively refine each element of the top-level curve $P_{\text{top}}$ (i.e., a multiscale node in the corresponding quadtree/octree)---at each level, a minimum Hamiltonian path is found with our flexible Hamiltonian path method that improves the Hamiltonian path method on grid graphs~\cite{Itai:JOC1982,Mitchell:Jres05}~(\autoref{fig:multiscaleSteps}~(c)). 
	Finally, the linearization is achieved: both the data value and the scale of the vertex are recorded
	(\autoref{fig:multiscaleSteps}~(d)).
	
	The objective function is approximately minimized during the process.
	The data value coherency term is minimized approximately with the flexible Hamiltonian path generation for each level in a block, and by finding the best entry node during adaptive refinement; the locality term is implicitly minimized by the hierarchical block-by-block advancing of the curve similar to the Peano-Hilbert curve.
	
	In the rest of this section, we explain the flexible Hamiltonian path generation method, and then, the adaptive refinement process; finally, we discuss scenarios when the multiscale technique or the regular grids technique should be used.
	
	\subsection{Flexible Hamiltonian Path for 2D and 3D Grid Graphs}
	\label{sec:hamPath}
	Typical Hamiltonian path methods solve the problem, i.e., $(G, v_s, v_t)$, on a regular grid $G$ with distinct, explicitly given entry and exit vertices $v_s$ and $v_t$. 
	However, this is not appropriate for our method as the adjacent vertices are of different scales.
	\rev{For example, as shown in~\autoref{fig:multiscaleSteps}, the exit vertex of the top-level block 3 (\autoref{fig:multiscaleSteps}~(b)) cannot be known beforehand, but only the exit face of the block is known given the top-level space-filling curve.}
	Therefore, in our formulation, we rewrite the Hamiltonian path problem as
	\begin{equation}
	(G, v_s, F_t)\;,
	\end{equation}
	\rev{where $F_t$ is the exit side/face of the bounding rectangle/box of $G$.}
	The task is then to calculate the minimum path from $v_s$ to a valid vertex on $F_t$.
	
	We have to compute the minimum path among all possible paths from entry point $v_s$ to all valid vertices on $F_t$.
	Here, the objective function $f(P)$ is simplified to describe only the data value coherency of the sequence, and it is defined as the sum of gradient magnitude of data values $s(V)$ along the path:
	\begin{equation}
	f(P) = \sum_{i = 1}^{n-1}||s(v_{i+1}) - s(v_{i})||\;.
	\end{equation}
	Therefore, a smoothly changing path is encouraged and a path that fluctuates significantly is punished.

\begin{figure*}[tb]
		\centering
		\includegraphics[width = 0.9\linewidth]{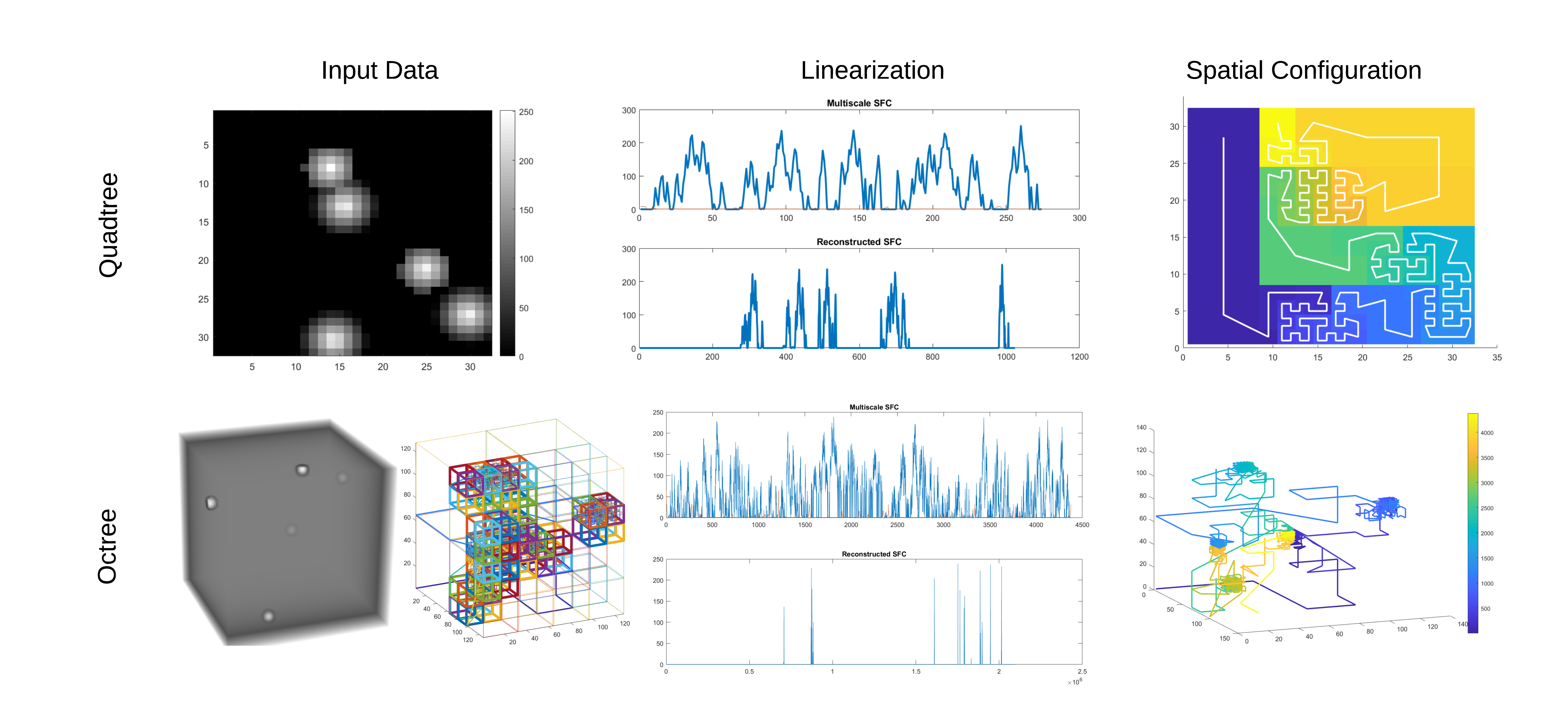}\vspace{-2em}
		\caption{Data-driven space-filling curves for quadtree and octree. The input data are shown in the first column, the linearizations in the second column, and the spatial configurations of the space-filling curves in the third column. }\vspace{-1em}
		\label{fig:quadtree}
	\end{figure*}

	\unsure{
		We can find (or show the nonexistence of) a Hamiltonian path through given entry and exit vertices for small grids directly by using exhaustive search.}
	A larger graph has to be partitioned into smaller ones: in practice, the largest graph that can be solved directly is $8\times4$ for 2D or $4\times4\times2$ for 3D on our test machines.
	The limitation of the partitioning is that it may break coherent features in space, e.g., in \autoref{fig:quadtree}, single disks/spheres are occasionally broken into different blocks and become less coherent compared to being in the same block. Therefore, we suggest as few partitions as possible if it is supported by the hardware and computational time allows.  
	\rev{The partition is based on the relationship between the entry face and exit face of the bounding box/rectangle of the graph.} 
	
\begin{figure}[htb]
		\centering
		\includegraphics[width = 0.9\linewidth]{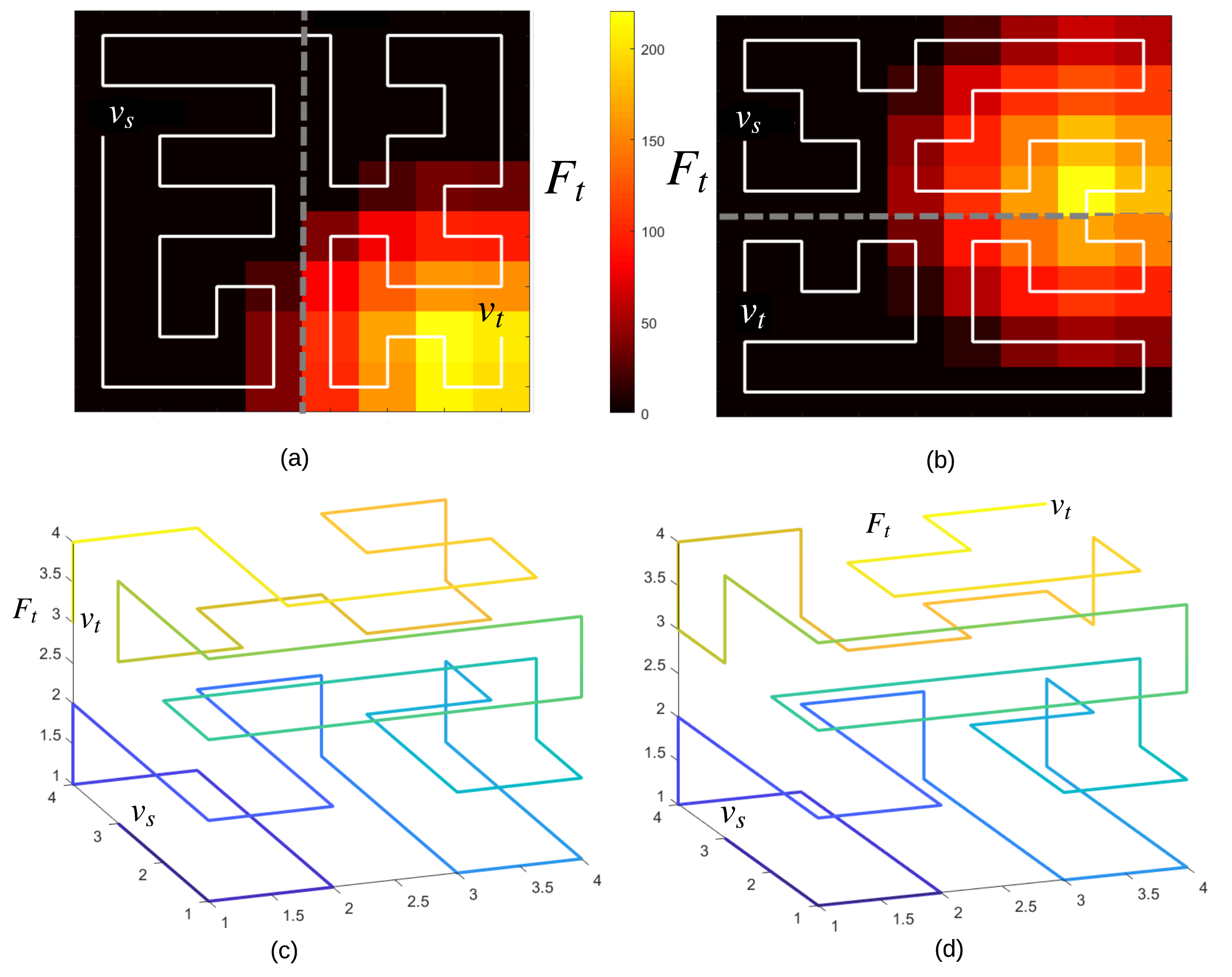}\vspace{-1em}
		\caption{Flexible Hamiltonian paths for 2D and 3D grid graphs. Exit sides are on the (a) right and (b) left for these examples of 2D graphs. The example 3D graph has exit faces on the (c) left and at the (d) top.}\vspace{-1em}
		\label{fig:hamPath}
	\end{figure}

	Examples of our flexible Hamiltonian path technique are shown in \autoref{fig:hamPath}: a horizontal partitioning and a vertical partitioning of 2D graphs are shown in~\autoref{fig:hamPath}~(a) and (b), respectively; exit faces $F_t$ on the left and top for 3D graphs are shown in (c) and (d). 
	Since the flexible Hamiltonian path method is the building block of our multiscale space-filling curve techniques, exhaustive search is implemented in an non-recursive fashion using stacks to improve efficiency.

	\subsection{Adaptive Refinement}
	The refinement method \textbf{refine}---as described in Algorithm~\autoref{alg:quadtreeRefine}---is the core of the space-filling curve for multiscale data.
	If any of the nodes within the block is not a leaf node, it has to be refined all the way down to the finest level in a data-driven fashion. 
	The key is to determine the suitable entry node for blocks at different levels (the \textbf{findBestEntry} function in Algorithm~\autoref{alg:quadtreeRefine}): we keep track of the last vertex $v_{\text{last}}$ in the Hamiltonian path and utilize it to find the matching entry node in the next block, i.e., the node within the adjacent block to $v_{\text{last}}$ that has the minimum difference to its data value.
	The combination of this process and the Hamiltonian path generation function \textbf{linearizeHamPath} (Section~\ref{sec:hamPath}) approximates the minimization.
	
	\begin{scriptsize}
		\begin{algorithm}[htb]
			\caption{Refinement of a Multiscale Block}\label{alg:quadtreeRefine}
			\begin{algorithmic}[1]
				\Procedure{refine}{$\{I_1, I_2, \cdots, I_{L_c}\}$, $T$, block, $v_{\text{last}}$}
				\State $v_s$ $\gets$ \textbf{findBestEntry}($v_{\text{last}}$)	\Comment{finds the best entry vertex $v_s$ (data-driven) }	
				\If{\textbf{needRefine}(block)}
				\State  $P_{\text{curr}} \gets$ [] \Comment{$P_{\text{curr}}$---SFC of the current multiscale block}
				\State  $P_{\text{currTop}} \gets$ \textbf{linearizeHamPath}($I_{\text{block}}$, $T$, $v_s$)\Comment{finds the top-level SFC  $P_{\text{currTop}}$ of the current data block $I_{\text{block}}$ (data-driven)}	
				\For{$i$ in range(1, length($P_{\text{currTop}}$))}
				\State ctopBlock $\gets$ T($P_{\text{currTop}}[i]$) \Comment{ctopBlock: top-level sub-block within the current block}
				\If{\textbf{needRefine}(ctopBlock)}
				\If{$i\geq 2$}
				\State  $v_{\text{last}} \gets P_{\text{currTop}}[i-1]$
				\EndIf
				
				\State $P_{\text{finer}} \gets$ \textbf{refine}($\{I_1, I_2, \cdots, I_{L_c}\}$, $T$, ctopBlock, $v_{\text{last}}$)\Comment{recursively computes the SFC of ctopBlock}
				\State $P_{\text{curr}} \gets [P_{\text{curr}}, P_{\text{finer}}]$\Comment{appends $P_{\text{curr}}$ with $P_{\text{finer}}$}
				\Else
				\State $P_{\text{curr}} \gets [P_{\text{curr}},P_{\text{currTop}}[i]]$ \Comment{appends $P_{\text{curr}}$ with $P_{\text{currTop}}[i]$}
				\EndIf
				\EndFor
				\Else
				\State $P_{\text{curr}} \gets$ \textbf{linearizeHamPath}($I_{\text{block}}$, block, $v_s$)\Comment{data-driven }	
				
				\EndIf
				\State \textbf{return} $P_{\text{curr}}$
				\EndProcedure
			\end{algorithmic}
		\end{algorithm}
	\end{scriptsize}
	
	\autoref{fig:quadtree}~(Quadtree) shows the linearization with our data-driven technique for quadtree on a synthetic image (\autoref{fig:quadtree} (Quadtree, first column)).
	\rev{The resulting multiscale linearizations and their reconstructed linearizations are shown in the second column of~\autoref{fig:quadtree}~(row 1). 
		Here, ``reconstructed" refers to generating the linearization back to the regular grid using the visit order of the coordinates of multiscale nodes and their scale information.
	}
	It can be seen that our technique preserves the value signatures of five disks---each as a peak---which is more prominent in the reconstructed space-filling curve.  \autoref{fig:quadtree} (row 1, third column) shows the geometry of the space-filling curve over the quadtree color mapped by the traversal order (from blue to yellow).
	An octree data of five spheres is shown in~\autoref{fig:quadtree}~(Octree): the linearization using our technique~\autoref{fig:quadtree}~(Octree, second column, top) preserves the value pattern of five spheres, more evident in the reconstructed version~\autoref{fig:quadtree}~(Octree, second column, bottom). 
	The spatial configuration of the space-filling curve is shown in~\autoref{fig:quadtree}~(Octree, third column), with the color map showing the scan order.
	\section{Evaluation}
	\label{sec:eval}
	\begin{figure}[htb]
		\centering
		\begin{tabular}{p{0.05\linewidth}cc}
			&Data value & Radial Euclidean distance\\
			[-2ex]
			\rot{90}{~~~~~~~~\small{2D}}&\hspace{-2em}
			\subfloat[]{\includegraphics[width=0.45\linewidth]{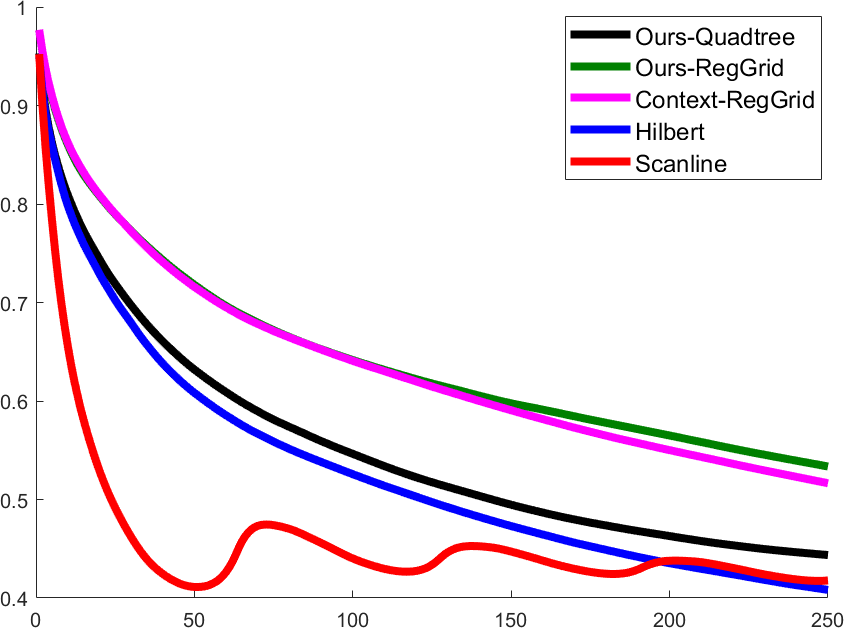}}&\hspace{-1em}
			\subfloat[]{\includegraphics[width=0.45\linewidth]{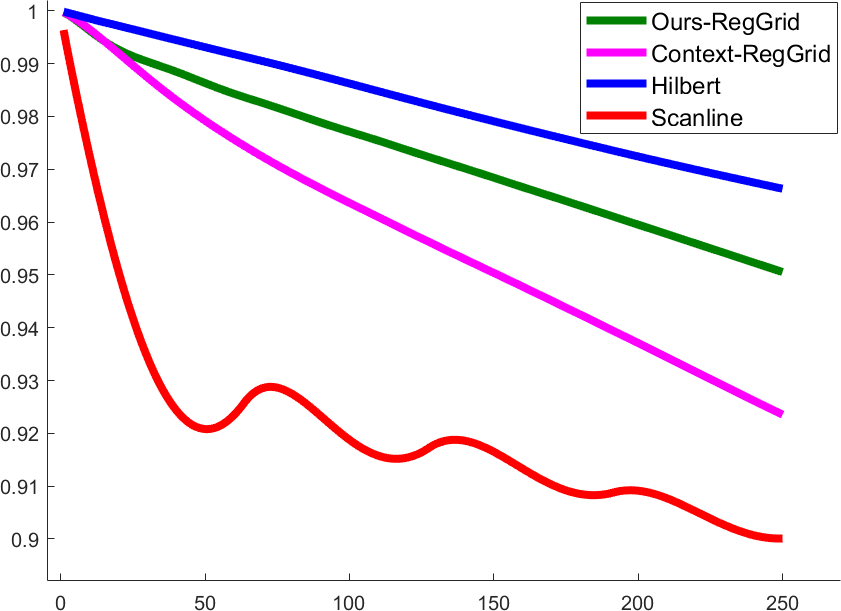}}\\
			[-2ex]
			\rot{90}{~~~~~~~~\small{3D}}&\hspace{-2em}
			\subfloat[]{\includegraphics[width=0.45\linewidth]{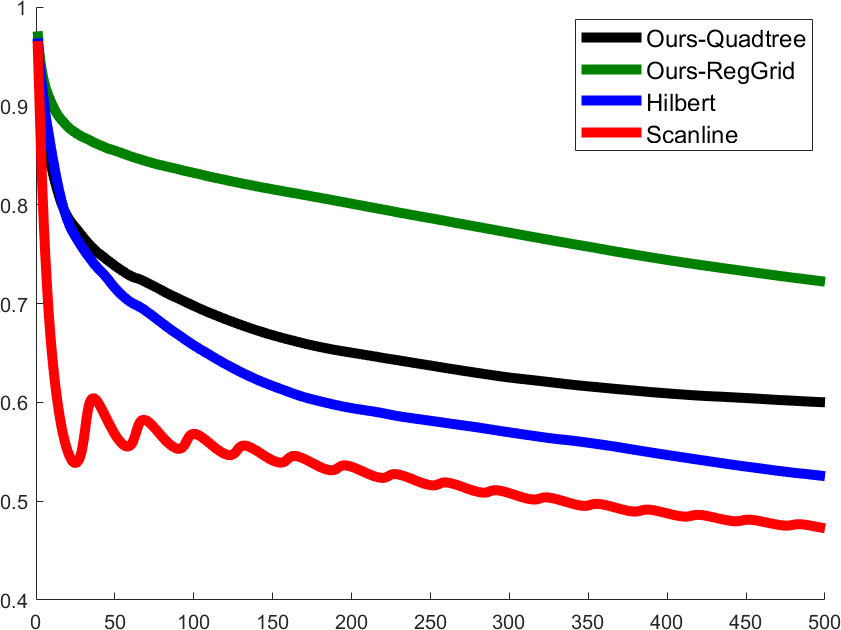}}&\hspace{-1em}
			\subfloat[]{\includegraphics[width=0.45\linewidth]{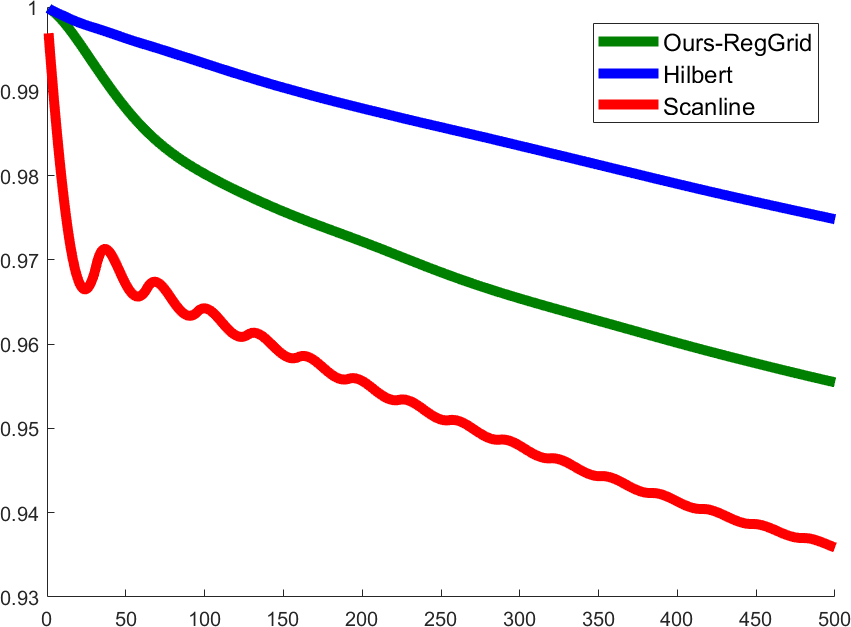}}
		\end{tabular}\vspace{-0.5em}
		\caption{Autocorrelations of data value (first column) and radial distance (second column) for our 2D techniques (first row) and our 3D techniques (second row). Note that larger autocorrelation means better coherency.}\vspace{-1em}
		\label{fig:autoCorr}
	\end{figure}
	
	\begin{figure*}[htb]
		\centering
		\includegraphics[width = 0.92\linewidth]{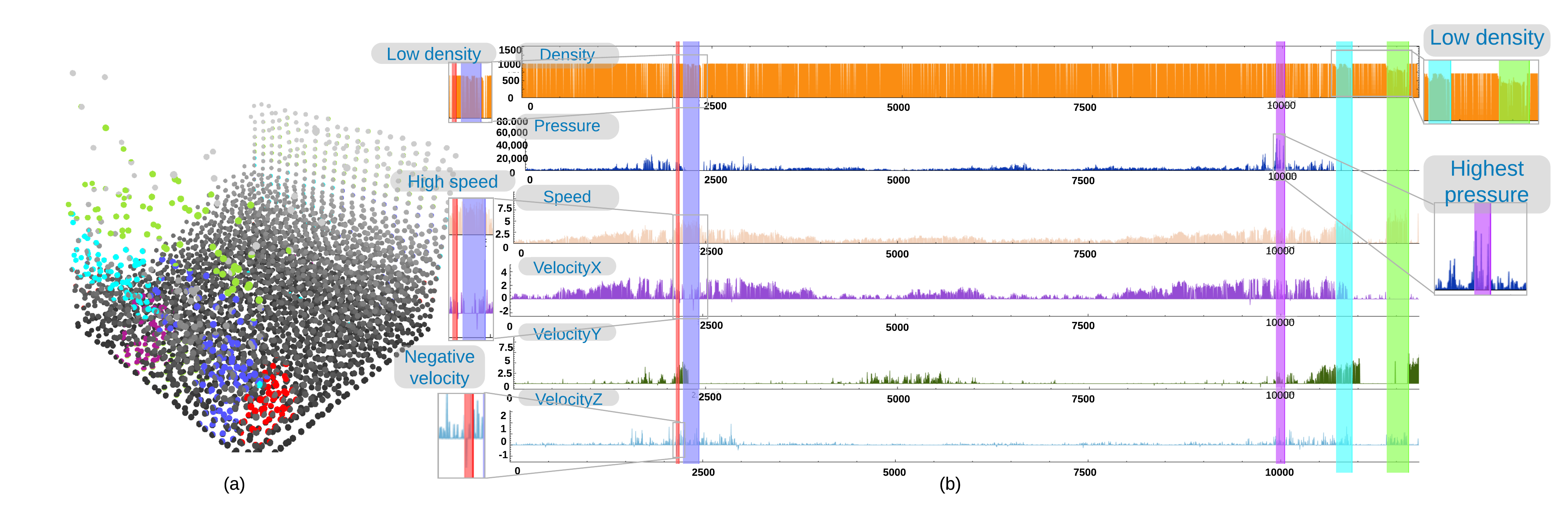}\vspace{-1.5em}
		\caption{Visualization of an SPH simulation of dam break: (a) rendering of particles with brushed regions, (b) multivariate line charts generated using our octree-based data-driven space-filling curve.}\vspace{-1.5em}
		\label{fig:sph}
	\end{figure*}
	Our method is evaluated by numerical comparison of autocorrelations of our techniques ($\alpha = 0.1$ for regular grid techniques) to existing linearization methods.
	Autocorrelation is the correlation of a signal and a shifted copy of the signal; the measurement indicates the coherency of a signal, and is suitable for measuring the effectiveness of space-filling curves~\cite{Dafner:cgf00}. 
	In our evaluation, autocorrelations of two measures are calculated: 1) autocorrelation of linearized data values $u(i)$ that measure the data coherency of space-filling curves; 2) autocorrelation of radial Euclidean distances of elements in the linearization $t(i)$ that measures the spatial coherency, i.e., locality, of the curves. 
	The definitions of the two measures are shown as follows: 
	\begin{equation}
	u(i) = s(P(i))\;,\qquad
	t(i) = ||[P(i).x, P(i).y, P(i).z] - [0,0,0]||\;.\nonumber
	\end{equation}
	\rev{Note that the distance measure $t(i)$ is applicable only to regular grids and $P(i).z=0$ for 2D cases.}
	
	We use benchmark datasets commonly employed in scientific visualization and average the autocorrelations of each dataset for each linearization technique.
	Specifically, 11 datasets---typically, slices of volume data (one slice each of downsampled volume datasets of aneurysm, beetle, bonsai, MRI brain, engine, foot, fuel, hurricane Isabel, neghip, and nucleon---all from a public volume data library\footnote{\url{http://schorsch.efi.fh-nuernberg.de/data/volume/}}; and an image of 5 randomly placed disks)---are used in the evaluation of 2D methods, and 5 volumetric datasets (fuel, neghip, nucleon, heart ischemia, and a procedural volume generated with a tangle function; all downsampled to $32^3$) are used for 3D. 
	Autocorrelations are shown in~\autoref{fig:autoCorr}, where the horizontal axis is the lag (shift) of the signal, and the vertical axis is the value of normalized autocorrelation. 
	
	Averaged autocorrelations of data values in 2D (\autoref{fig:autoCorr}~(a)) indicate that our regular grid method (green) has almost the same feature coherency as the context-based curve (purple) as they are overlapping, and both perform much better than the Peno-Hilbert curve (blue) and the scanline (red); our data-driven method for quadtree (black) performs better than the Peano-Hilbert curve and scanline. 
	In terms of averaged autocorrelations of radial distance (\autoref{fig:autoCorr}~(b)), the Peano-Hilbert curve (blue) performs best and is followed by our data-driven method (green), and then the context-based method (purple); the scanline method (red) has much worse performances than other techniques.
	
	For 3D data, the evaluation compares our regular grid-based data-driven technique, our multiscale technique for octree, the Peano-Hilbert curve, and the scanline.
	As shown in~\autoref{fig:autoCorr}~(c), our regular-grid method (green) tops other techniques for averaged autocorrelation of data value, and our octree technique (black) follows, and then, the Peano-Hilbert curve (blue), and the scanline (red).
	For autocorrelations of radial distance (\autoref{fig:autoCorr}~(d)), our regular grid method is better than the scanline but out-performed by the Peano-Hilbert curve. 
	
	The evaluation confirms that our data-driven technique balances the data value coherency and locality coherency, and is more flexible than existing techniques.
	\rev{The comparisons also suggest that our regular grid techniques have better data value coherency performance than our multiscale techniques---the former is preferred when high-quality linearization is needed for volumetric data and computational time is not a limiting factor.}
	
	\begin{figure*}[htb]
		\centering
		\subfloat{\includegraphics[width = 0.76\linewidth]{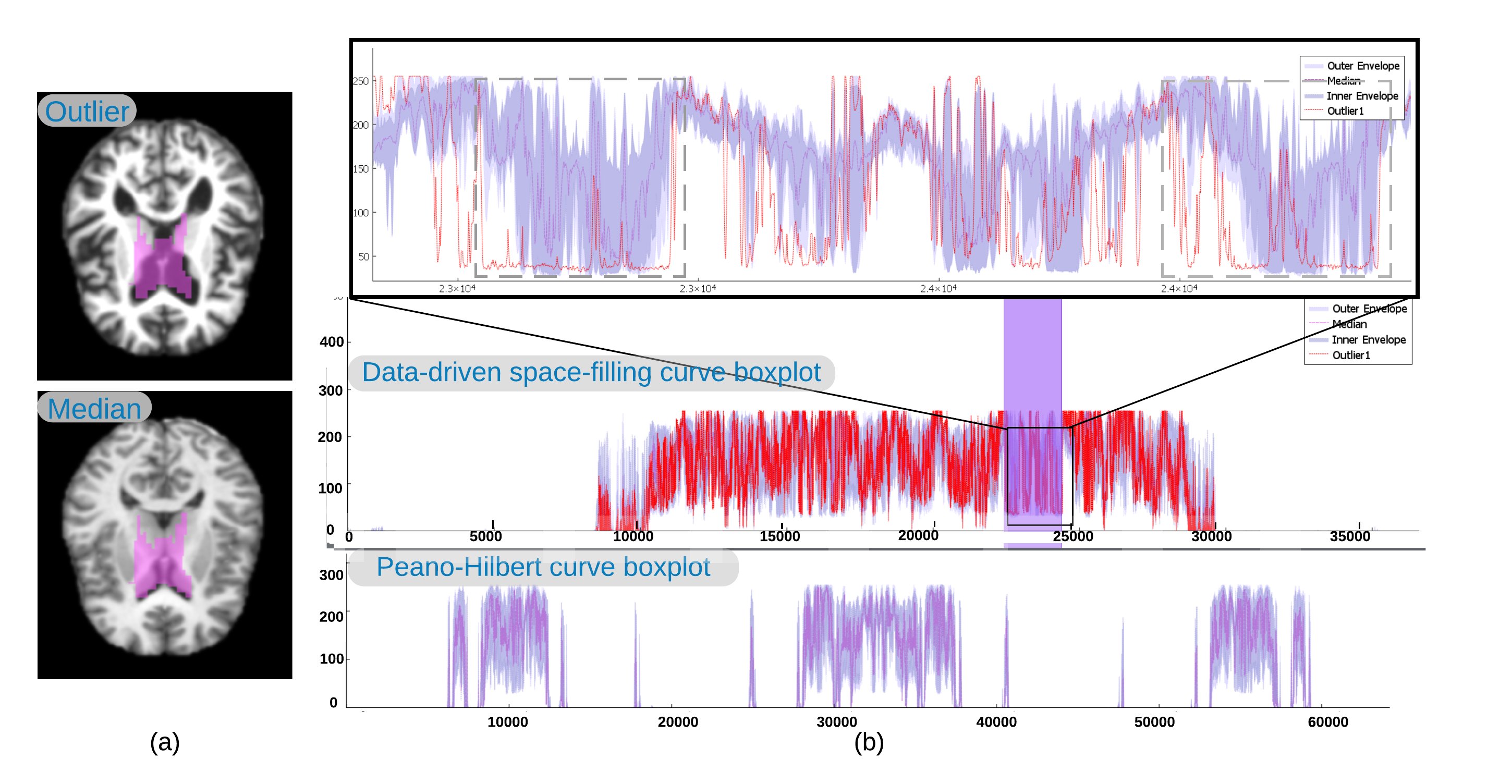}}\vspace{-1.5em}
		\caption{Visualization of a brain atlas with our data-driven space-filling curve and a Peano-Hilbert curve (with images padded to size of 256 $\times$ 256 with zeros). The median image (bottom) and an outlier image (top) are shown in (a) with brushing-and-linking (purple box) on the space-filling curves (b, center). The 1D layout with the space-filling curve allows for easy interaction, including brushing and zooming on spatial details, and it supports rendering boxplots that separate the brain from its surrounding and show that the outlier has wider low-value regions than the band at the lateral ventricle (red curves in the gray boxes in the zoom-in). }\vspace{-0.5em}
		\label{fig:oasis}
		
		\subfloat{\includegraphics[width=0.95\linewidth]{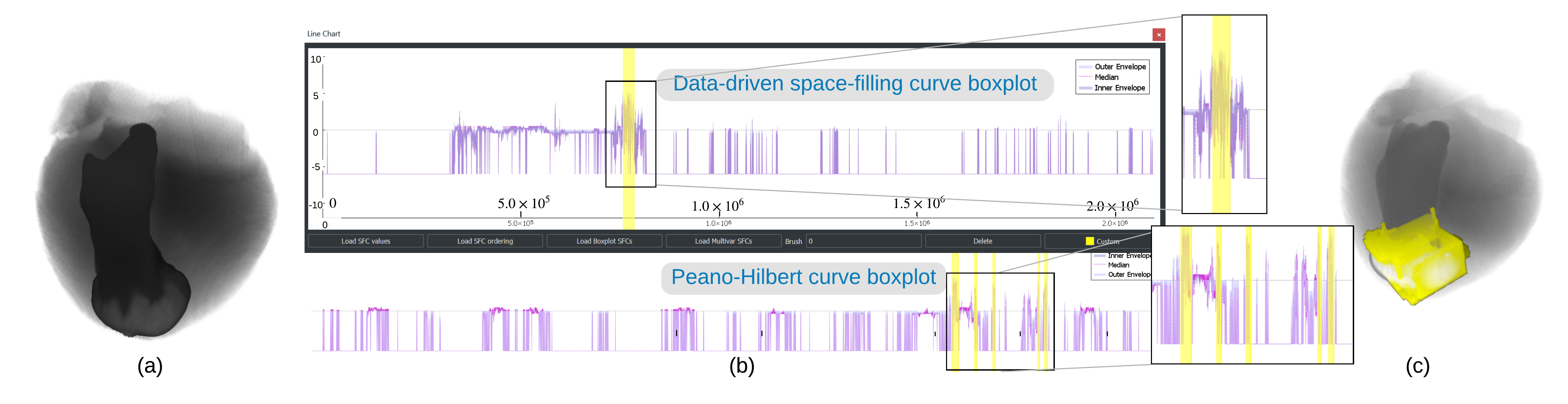}}\vspace{-1.5em}
		\caption{Ensemble visualizations of a heart ischemia simulation. The median member is volume-rendered in (a)---with 1D functional boxplots linearized with (b) our data-driven space-filling curve (top) and a Peano-Hilbert curve (bottom). The ischemic region that has potential value greater than 3~eV is selected in the boxplots and highlighted in (c).}\vspace{-1em}
		\label{fig:heart}
	\end{figure*}
	
	Our multiscale techniques are most suitable for multiscale data by nature, e.g., multiscale simulation of particles.
	An octree can be built to have a few (even just one) particles in the finest level node so that accurate data values of almost all particles could be preserved.
	However, for regular grids, this is more difficult if not impossible.
	Either more particles are averaged out or a very fine grid has to be built.
	In addition, multiscale techniques are faster than regular grid techniques as fewer nodes have to be visited in multiscale structures compared to regular grids for the same input data.
	
	\rev{In contrast, our regular grid techniques yield more coherent linearization results than our multiscale curves.}
	This is due to that the multiscale curve uses the top-down approach to ensure a Hamiltonian path exists, but it breaks coherent features in space on certain occasions as demonstrated in~\autoref{fig:quadtree}~(Octree).
	\rev{The aforementioned numerical comparison of coherency confirms that the regular grid techniques have higher coherency than multiscale techniques.}
	
	\rev{Therefore, we recommend using the multiscale technique for intrinsically multiscale data, especially, point datasets, and for reducing computation time.
		The regular grids techniques are recommended for higher linearization quality for images and volumetric data on uniform grids.
		Furthermore, preprocessing the input data with a segmentation could improve the coherency, and, potentially, the efficiency of our method (for regular grids, fewer comparisons are needed when neighbors are homogeneous).
	}
	
	\section{Visualization, User Interaction, and Implementation}
	\label{sec:visandui}
	Our method facilitates visualizations that use the horizontal axis for any spatial configuration along the space-filling curve, freeing the vertical space to visualize values aligned within the spatial configuration.
	For multivariate data, each variable can be visualized as a line plot (\autoref{fig:sph}), whereas for ensemble data, functional boxplots are used (\autoref{fig:teaser},\ref{fig:oasis},\ref{fig:heart}).
	Here, we employ the surface boxplot~\cite{SCI:Gen2014a} for 2D ensemble datasets; an extension of the method~\cite{SCI:Gen2014a} to 3D is applied to 3D ensemble datasets.
	The conventional color scheme used for boxplots is adopted. 
	
	We have built an interactive visualization tool to support the exploration of space-filling curve-based visualization of datasets. 
	The tool comprises three linked views: a line plot view that shows the linearized data; a 3D renderer that renders volumetric data with direct volume rendering, and particle datasets with polygon-based rendering; a 2D renderer that shows the data slice.
	The line plot view is linked with 2D and 3D renderers using the scan order of the space-filling curve that records the pixel ID of the line plot and its associated pixels/voxels in the original data.
	Brushing-and-linking allows us to brush regions in the line plot view, highlighting them in the 2D and 3D renderings.
	Interactive zooming and panning are supported in the line plot view so that both the overall structure and details of the line visualizations can be examined.
	
	Our space-filling curve techniques were implemented using Matlab.
	The visualization tool is built using C++, Qt, and OpenGL, and is accelerated by the GPU. 
	The QCustomPlot library~\cite{Qcustomplot} was used to aid the implementation of the line plot view. 
	Our method was tested on a 2019 13-inch Macbook Pro with 2.3 GHz Intel i5 CPU, 8 GB main memory, and an Intel Iris 655 integrated GPU.
	The data-driven space-filling curve only needs to be computed once for a given dataset, and the computation time depends on the number of vertices in the graph representation of the dataset. 
	\rev{Timings of generating data-driven space-filling curves for examples of the paper are summarized in Table~\ref{tbl:timing}.
		Full interactivity was achieved for the exploration of all examples.}
	\begin{table}[htb]
		\centering
		\caption{Computation time of data driven space-filling curves on example datasets.}
		\footnotesize
		\begin{tabular}{ccc}
			Dataset &
			Size	&
			Time	\\
			\toprule
			Nucleon slice & 64$\times$64 pixels& 12s\\
			Nucleon &  32$\times$32$\times$32 voxels & 24s \\
			Brain atlas & 176$\times$208 pixels& 3m39s\\
			SPH & 4000 particles/11796 octree nodes & 43s \\
			Myocardial ischemia & 128$\times$128$\times$128 voxels& 4h31m \\
			\bottomrule
		\end{tabular}
		\label{tbl:timing}
	\end{table}
	
	\section{Examples}
	\label{sec:examples}
	We demonstrate the usefulness of our method with examples of multiscale multivariate particle data, ensemble of medical images on 2D regular grids, and volumetric ensemble datasets on regular grids.
	\unsure{Visualizations of ensemble datasets are based on linearizations of the median members.}
	The smooth particle hydrodynamics (SPH) dataset shown in~\autoref{fig:sph} is a timestep in a dam break simulation~\cite{Reinhardt:IV17}; the dataset contains particles with six attributes: density, pressure, speed, and velocity in $X$, $Y$, and $Z$ directions, respectively. 
	The data is decomposed into an octree and linearized using our octree-based data-driven curves.
	\unsure{Data values of all attributes are linearized with the spatial layout of the space-filling curve of the pressure attribute.}
	We highlight regions that are distinct from their neighborhood in the linearizations: low values of density, high values of pressure, and high values of speed and velocity. 
	Here, the most prominent feature is the highest pressure region (brushed in purple) with values over 50,000 as shown in the zoom-in.
	The particles within the brushed regions can be seen in the 3D rendering (\autoref{fig:sph}~(a)).  
	Our method yields a new visual debugging method that shows clear, non-occluded quantities of each attribute embedded in a spatial context. 
	It could complement non-spatial multivariate plots~\cite{Reinhardt:IV17} for a more comprehensive visual debugging system. 
	
	\autoref{fig:oasis} shows the visualization of a series of open-access MRI slices~\cite{Marcus:oasis}. 
	The boxplot linearized with our method (\autoref{fig:oasis}~(b, center)) exhibits a more coherent feature that is more concentrated than in the Peano-Hilbert curve linearization (\autoref{fig:oasis}~(b, bottom)).
	As shown in the zoom-in of~\autoref{fig:oasis}~(b, top), the outlier (the red curve) has wider low-value regions (inside the gray boxes) than the band as shown in the zoom-in.
	With brushing-and-linking, it is confirmed that the outlier image (\autoref{fig:oasis}~(a, top)) has larger lateral ventricle area than the median image (\autoref{fig:oasis}~(a, bottom)) with the brush on the boxplot linearized with our data-driven space-filling curve.
	
	The myocardial ischemia dataset was generated by image-based, experimentally derived cardiac electrical potential simulation~\cite{SCI:Bur2018a}.
	\rev{We use a subset of ensemble runs of the simulation and sample the data on regular grids for our experiment.}
	Here, we are interested in the acute ischemic regions associated with mean potentials greater or equal to 3~eV. 
	As shown in~\autoref{fig:heart}~(b, top), the linearized 3D boxplot using data-driven technique yields more concentrated global features than the linearization with the Peano-Hilbert curve (\autoref{fig:heart}~(b, bottom)). The region of interest (high potential regions) is bounded in a small neighborhood with our method that could be selected with one brush (\autoref{fig:heart}~(b, top)), whereas the Peano-Hilbert curve yields a more scattered result---a large number of brushes are required (\autoref{fig:heart}~(b, bottom)).
	The volume rendering (\autoref{fig:heart}(a)) of the median ensemble member shows that the region of interest is spatially continuous (white in the rendering); the highlighted regions in space (\autoref{fig:heart}~(c)) verify that our method gives good coherency of the feature.

	\section{Conclusion and Future Work}
	
	We have introduced data-driven space-filling curves for 2D and 3D visualization.
	We have designed our methods to preserve coherency of both data value and locality after the mapping from the spatial domain to 1D.
	The methods are applicable for data on regular grids and in multiscale. 
	\rev{We have modeled the problem as finding a Hamiltonian path that approximates the minimum of an objective function that blends a data value term and a locality term.}
	Two variants of techniques are available for regular grids and multiscale data (quadtrees and octrees).
	The effectiveness of our method has been evaluated by comparing to existing methods on various datasets with qualitative visual comparison and quantitative comparisons of autocorrelations.
	We have confirmed that existing methods cannot preserve both data features and locality after linearization.
	Through multivariate and ensemble visualization examples with a wide range of real-world datasets, we have demonstrated the usefulness of our data-driven space-filling curves. 
	
	In the future, we would like to extend our method for time-varying data to understand the coherency in time.
	\rev{The positional term for regular grids requires uniform blocks of a user-defined size, which should be improved to be data-driven.}
	The method could be used for multi-field visualization such that different field data, e.g., scalar, vector, and tensor, could be visualized in a linear layout for non-occluded comparisons and investigation of correlations. 
	\rev{Finally, we would also like to accelerate our method with parallel computing to support larger datasets.}
	
	\acknowledgments{
		This work is supported in part by the Intel Graphics and Visualization Institutes of XeLLENCE, the National Institute of General Medical Sciences of the National Institutes of Health under grant numbers P41 GM103545 and R24 GM136986 and the Department of Energy under grant number DE-FE0031880, and by the Deutsche Forschungsgemeinschaft (DFG, German Research Foundation) Project-ID 251654672 -- TRR 161 (project B01). The GPU used for this research was donated by the Nvidia Corporation. The SPH dataset is provided by Stefan Reinhardt. 
	}

	\bibliographystyle{abbrv-doi}

	\bibliography{spaceFillingCurves}

\begin{thebibliography}{10}

\bibitem{Bollobas:1979}
B.~Bollobas.
\newblock {\em Graph Theory: An Introductory Course}.
\newblock Springer-Verlag, New York, 1979. doi: {{%
10\hspace{.1pt}\discretionary{.}{%
}{.}\hspace{.4pt}1007\discretionary{/}{%
}{/}978\discretionary{%
}{-}{-}1\discretionary{%
}{-}{-}4612\discretionary{%
}{-}{-}9967\discretionary{%
}{-}{-}7}}


\bibitem{Briais:CHES2012}
S.~Briais, S.~Caron, J.-M. Cioranesco, J.-L. Danger, S.~Guilley, J.-H. Jourdan,
  A.~Milchior, D.~Naccache, and T.~Porteboeuf.
\newblock {3D} hardware canaries.
\newblock In E.~Prouff and P.~Schaumont, eds., {\em Cryptographic Hardware and
  Embedded Systems -- CHES 2012}, pp. 1--22. Springer, Berlin, Heidelberg,
  2012.

\bibitem{SCI:Bur2018a}
B.~Burton, K.~Aras, W.~Good, J.~Tate, B.~Zenger, and R.~MacLeod.
\newblock A framework for image-based modeling of acute myocardial ischemia
  using intramurally recorded extracellular potentials.
\newblock {\em Annals of Biomedical Engineering}, 2018. doi: {{%
10\hspace{.1pt}\discretionary{.}{%
}{.}\hspace{.4pt}1007\discretionary{/}{%
}{/}s10439\discretionary{%
}{-}{-}018\discretionary{%
}{-}{-}2048\discretionary{%
}{-}{-}0}}


\bibitem{campbell03a}
P.~M. Campbell, K.~D. Devine, J.~E. Flaherty, L.~G. Gervasio, and J.~D.
  Teresco.
\newblock Dynamic octree load balancing using space-filling curves.
\newblock Technical Report CS-03-01, Williams College Department of Computer
  Science, 2003.

\bibitem{Dafner:cgf00}
R.~Dafner, D.~Cohen-Or, and Y.~Matias.
\newblock Context-based space filling curves.
\newblock {\em Computer Graphics Forum}, 19(3):209--218, 2000. doi: {{%
10\hspace{.1pt}\discretionary{.}{%
}{.}\hspace{.4pt}1111\discretionary{/}{%
}{/}1467\discretionary{%
}{-}{-}8659\hspace{.1pt}\discretionary{.}{%
}{.}\hspace{.4pt}00413}}


\bibitem{Demir:VIS14}
I.~{Demir}, C.~{Dick}, and R.~{Westermann}.
\newblock Multi-charts for comparative {3D} ensemble visualization.
\newblock {\em IEEE Transactions on Visualization and Computer Graphics},
  20(12):2694--2703, 2014. doi: {{%
10\hspace{.1pt}\discretionary{.}{%
}{.}\hspace{.4pt}1109\discretionary{/}{%
}{/}TVCG\hspace{.1pt}\discretionary{.}{%
}{.}\hspace{.4pt}2014\hspace{.1pt}\discretionary{.}{%
}{.}\hspace{.4pt}2346448}}


\bibitem{Qcustomplot}
E.~Eichhammer.
\newblock {Qt Plotting Widget QCustomPlot}.
\newblock {https://www.qcustomplot.com/index.php/introduction}, 2018.

\bibitem{Faloutsos:TSE88}
C.~Faloutsos.
\newblock Gray codes for partial match and range queries.
\newblock {\em IEEE Transactions on Software Engineering}, 14(10):1381–1393,
  1988. doi: {{%
10\hspace{.1pt}\discretionary{.}{%
}{.}\hspace{.4pt}1109\discretionary{/}{%
}{/}32\hspace{.1pt}\discretionary{.}{%
}{.}\hspace{.4pt}6184}}


\bibitem{SCI:Gen2014a}
M.~Genton, C.~Johnson, K.~Potter, G.~Stenchikov, and Y.~Sun.
\newblock Surface boxplots.
\newblock {\em Statistical Journal}, 3(1):1--11, 2014.

\bibitem{Hilbert:1891}
D.~Hilbert.
\newblock {Ueber die stetige Abbildung einer Line auf ein
  Fl{\"a}chenst{\"u}ck}.
\newblock {\em Mathematische Annalen}, 38(3):459--460, 1891. doi: {{%
10\hspace{.1pt}\discretionary{.}{%
}{.}\hspace{.4pt}1007\discretionary{/}{%
}{/}BF01199431}}


\bibitem{Itai:JOC1982}
A.~Itai, C.~H. Papadimitriou, and J.~L. Szwarcfiter.
\newblock Hamilton paths in grid graphs.
\newblock {\em SIAM Journal on Computing}, 11(4):676--686, 1982. doi: {{%
10\hspace{.1pt}\discretionary{.}{%
}{.}\hspace{.4pt}1137\discretionary{/}{%
}{/}0211056}}


\bibitem{Marcus:oasis}
D.~S. Marcus, T.~H. Wang, J.~Parker, J.~G. Csernansky, J.~C. Morris, and R.~L.
  Buckner.
\newblock Open access series of imaging studies {(OASIS)}: Cross-sectional
  {MRI} data in young, middle aged, nondemented, and demented older adults.
\newblock {\em Journal of Cognitive Neuroscience}, 19(9):1498--1507, 2007. doi:
  {{%
10\hspace{.1pt}\discretionary{.}{%
}{.}\hspace{.4pt}1162\discretionary{/}{%
}{/}jocn\hspace{.1pt}\discretionary{.}{%
}{.}\hspace{.4pt}2007\hspace{.1pt}\discretionary{.}{%
}{.}\hspace{.4pt}19\hspace{.1pt}\discretionary{.}{%
}{.}\hspace{.4pt}9\hspace{.1pt}\discretionary{.}{%
}{.}\hspace{.4pt}1498}}


\bibitem{Matias:spacefillcurve}
Y.~Matias and A.~Shamir.
\newblock A video scrambling technique based on space filling curves (extended
  abstract).
\newblock In {\em Advances in Cryptology --- CRYPTO '87}, pp. 398--417, 1988.
  doi: {{%
10\hspace{.1pt}\discretionary{.}{%
}{.}\hspace{.4pt}1007\discretionary{/}{%
}{/}3\discretionary{%
}{-}{-}540\discretionary{%
}{-}{-}48184\discretionary{%
}{-}{-}2\_35}}


\bibitem{SCI:Mir14b}
M.~{Mirzargar}, R.~T. {Whitaker}, and R.~M. {Kirby}.
\newblock Curve boxplot: Generalization of boxplot for ensembles of curves.
\newblock {\em IEEE Transactions on Visualization and Computer Graphics},
  20(12):2654--2663, 2014. doi: {{%
10\hspace{.1pt}\discretionary{.}{%
}{.}\hspace{.4pt}1109\discretionary{/}{%
}{/}TVCG\hspace{.1pt}\discretionary{.}{%
}{.}\hspace{.4pt}2014\hspace{.1pt}\discretionary{.}{%
}{.}\hspace{.4pt}2346455}}


\bibitem{Mitchell:Jres05}
W.~F. Mitchell.
\newblock Hamiltonian paths through two- and three-dimensional grids.
\newblock {\em Journal of Research of the National Institute of Standards and
  Technology}, 110(2):127--136, 2005. doi: {{%
10\hspace{.1pt}\discretionary{.}{%
}{.}\hspace{.4pt}6028\discretionary{/}{%
}{/}jres\hspace{.1pt}\discretionary{.}{%
}{.}\hspace{.4pt}110\hspace{.1pt}\discretionary{.}{%
}{.}\hspace{.4pt}012}}


\bibitem{Netzel:etvis16}
R.~{Netzel} and D.~{Weiskopf}.
\newblock Hilbert attention maps for visualizing spatiotemporal gaze data.
\newblock In {\em Second Workshop on Eye Tracking and Visualization (ETVIS)},
  pp. 21--25, 2016. doi: {{%
10\hspace{.1pt}\discretionary{.}{%
}{.}\hspace{.4pt}1109\discretionary{/}{%
}{/}ETVIS\hspace{.1pt}\discretionary{.}{%
}{.}\hspace{.4pt}2016\hspace{.1pt}\discretionary{.}{%
}{.}\hspace{.4pt}7851160}}


\bibitem{DBLP:journals/tvcg/ObermaierBJ16}
H.~Obermaier, K.~Bensema, and K.~I. Joy.
\newblock Visual trends analysis in time-varying ensembles.
\newblock {\em {IEEE} Transactions on Visualization and Computer Graphics},
  22(10):2331--2342, 2016. doi: {{%
10\hspace{.1pt}\discretionary{.}{%
}{.}\hspace{.4pt}1109\discretionary{/}{%
}{/}TVCG\hspace{.1pt}\discretionary{.}{%
}{.}\hspace{.4pt}2015\hspace{.1pt}\discretionary{.}{%
}{.}\hspace{.4pt}2507592}}


\bibitem{DBLP:journals/cga/ObermaierJ14}
H.~Obermaier and K.~I. Joy.
\newblock Future challenges for ensemble visualization.
\newblock {\em {IEEE} Computer Graphics and Applications}, 34(3):8--11, 2014.
  doi: {{%
10\hspace{.1pt}\discretionary{.}{%
}{.}\hspace{.4pt}1109\discretionary{/}{%
}{/}MCG\hspace{.1pt}\discretionary{.}{%
}{.}\hspace{.4pt}2014\hspace{.1pt}\discretionary{.}{%
}{.}\hspace{.4pt}52}}


\bibitem{Peano:1890}
G.~Peano.
\newblock Sur une courbe, qui remplit toute une aire plane.
\newblock {\em Mathematische Annalen}, 36(1):157--160, 1890. doi: {{%
10\hspace{.1pt}\discretionary{.}{%
}{.}\hspace{.4pt}1007\discretionary{/}{%
}{/}BF01199438}}


\bibitem{SCI:Pot2009b}
K.~Potter, A.~Wilson, P.-T. Bremer, D.~Williams, C.~Doutriaux, V.~Pascucci, and
  C.~Johnson.
\newblock {Ensemble-Vis}: A framework for the statistical visualization of
  ensemble data.
\newblock In {\em Proceedings of the 2009 {IEEE} International Conference on
  Data Mining Workshops}, pp. 233--240, 2009.

\bibitem{Raj:cga16}
M.~{Raj}, M.~{Mirzargar}, J.~S. {Preston}, R.~M. {Kirby}, and R.~T. {Whitaker}.
\newblock Evaluating shape alignment via ensemble visualization.
\newblock {\em IEEE Computer Graphics and Applications}, 36(3):60--71, 2016.
  doi: {{%
10\hspace{.1pt}\discretionary{.}{%
}{.}\hspace{.4pt}1109\discretionary{/}{%
}{/}MCG\hspace{.1pt}\discretionary{.}{%
}{.}\hspace{.4pt}2015\hspace{.1pt}\discretionary{.}{%
}{.}\hspace{.4pt}70}}


\bibitem{Reinhardt:IV17}
S.~{Reinhardt}, M.~{Huber}, O.~{Dumitrescu}, M.~{Krone}, B.~{Eberhardt}, and
  D.~{Weiskopf}.
\newblock {Visual Debugging of SPH Simulations}.
\newblock In {\em 21st International Conference Information Visualisation
  (IV)}, pp. 117--126, 2017. doi: {{%
10\hspace{.1pt}\discretionary{.}{%
}{.}\hspace{.4pt}1109\discretionary{/}{%
}{/}iV\hspace{.1pt}\discretionary{.}{%
}{.}\hspace{.4pt}2017\hspace{.1pt}\discretionary{.}{%
}{.}\hspace{.4pt}20}}


\bibitem{SCI:Ros2016a}
P.~Rosen, B.~Burton, K.~Potter, and C.~Johnson.
\newblock {muView}: A visual analysis system for exploring uncertainty in
  myocardial ischemia simulations.
\newblock In {\em Visualization in Medicine and Life Sciences III}, pp. 49--69.
  Springer Nature, 2016. doi: {{%
10\hspace{.1pt}\discretionary{.}{%
}{.}\hspace{.4pt}1007\discretionary{/}{%
}{/}978\discretionary{%
}{-}{-}3\discretionary{%
}{-}{-}319\discretionary{%
}{-}{-}24523\discretionary{%
}{-}{-}2\_3}}


\bibitem{Sagan:1994:SFC}
H.~Sagan.
\newblock {\em Space-Filling Curves}.
\newblock Springer-Verlag, New York, 1994. doi: {{%
10\hspace{.1pt}\discretionary{.}{%
}{.}\hspace{.4pt}1007\discretionary{/}{%
}{/}978\discretionary{%
}{-}{-}1\discretionary{%
}{-}{-}4612\discretionary{%
}{-}{-}0871\discretionary{%
}{-}{-}6}}


\bibitem{Sedgewick:algorithm}
R.~Sedgewick.
\newblock {\em Algorithms}.
\newblock Addison-Wesley Longman Publishing Co., Inc., Boston, USA, 1984.

\bibitem{SKUBALSKARAFAJLOWICZ19971305}
E.~Skubalska-Rafajłowicz.
\newblock Applications of the space-filling curves with data driven
  measure-preserving property.
\newblock {\em Nonlinear Analysis: Theory, Methods \& Applications},
  30(3):1305--1310, 1997. doi: {{%
10\hspace{.1pt}\discretionary{.}{%
}{.}\hspace{.4pt}1016\discretionary{/}{%
}{/}S0362\discretionary{%
}{-}{-}546X\discretionary{%
}{(}{(}97\discretionary{)}{%
}{)}00277\discretionary{%
}{-}{-}0}}


\bibitem{Sun:boxplots}
Y.~Sun and M.~G. Genton.
\newblock Functional boxplots.
\newblock {\em Journal of Computational and Graphical Statistics},
  20(2):316--334, 2011. doi: {{%
10\hspace{.1pt}\discretionary{.}{%
}{.}\hspace{.4pt}1198\discretionary{/}{%
}{/}jcgs\hspace{.1pt}\discretionary{.}{%
}{.}\hspace{.4pt}2011\hspace{.1pt}\discretionary{.}{%
}{.}\hspace{.4pt}09224}}


\bibitem{Weissenboeck:vis18}
J.~{Weissenb{\"o}ck}, B.~{Fr{\"o}hler}, E.~{Gr{\"o}ller}, J.~{Kastner}, and
  C.~{Heinzl}.
\newblock Dynamic volume lines: Visual comparison of {3D} volumes through
  space-filling curves.
\newblock {\em IEEE Transactions on Visualization and Computer Graphics},
  25(1):1040--1049, 2019. doi: {{%
10\hspace{.1pt}\discretionary{.}{%
}{.}\hspace{.4pt}1109\discretionary{/}{%
}{/}TVCG\hspace{.1pt}\discretionary{.}{%
}{.}\hspace{.4pt}2018\hspace{.1pt}\discretionary{.}{%
}{.}\hspace{.4pt}2864510}}


\bibitem{SCI:Mir2014a}
R.~T. {Whitaker}, M.~{Mirzargar}, and R.~M. {Kirby}.
\newblock Contour boxplots: A method for characterizing uncertainty in feature
  sets from simulation ensembles.
\newblock {\em IEEE Transactions on Visualization and Computer Graphics},
  19(12):2713--2722, 2013. doi: {{%
10\hspace{.1pt}\discretionary{.}{%
}{.}\hspace{.4pt}1109\discretionary{/}{%
}{/}TVCG\hspace{.1pt}\discretionary{.}{%
}{.}\hspace{.4pt}2013\hspace{.1pt}\discretionary{.}{%
}{.}\hspace{.4pt}143}}


\end{thebibliography}
\end{document}